\documentclass[longauth]{aa}

\usepackage{graphicx}
\usepackage{multirow}
\usepackage{tabularx}
\usepackage{lscape}
\usepackage{subfigure}
\usepackage{natbib}
\usepackage{url}
\newcommand{\cmark}{\ding{51}}%
\newcommand{\xmark}{\ding{55}}%
\usepackage[dvipsnames]{xcolor}
\usepackage{longtable}
\def \kms{km\,s$^{-1}$}
\def \ms{m\,s$^{-1}$}
\def \cms{cm\,s$^{-2}$}
\def \gcm{g~cm$^{-3}$}

\def \mearth{M$_\oplus$}
\def \mjup{M$_\mathrm{jup}$}

\def\sun{\odot}

\bibpunct{(}{)}{;}{a}{}{,} 

\usepackage{multirow}

\usepackage{amssymb}
\usepackage{pifont}
\usepackage{url}
\usepackage{threeparttable}
\usepackage{lipsum,multicol}
\usepackage[normalem]{ulem}

\def\approxinf{%
  \def\p{%
    \setbox0=\vbox{\hbox{$<$}}%
    \ht0=0.6ex \box0 }%
  \def\s{%
    \vbox{\hbox{$\sim$}}%
  }%
  \mathrel{\raisebox{0.7ex}{%
      \mbox{$\underset{\s}{\p}$}%
    }}%
}

\newcommand\methane{\ifmmode{{\rm CH}_{4}}\else{CH$_{4}$}\fi}
\newcommand\water{\ifmmode{{\rm H}_{2}{\rm O}}\else{H$_{2}$O}\fi}
\newcommand\carbdiox{\ifmmode{{\rm CO}_{2}}\else{CO$_{2}$}\fi}
\newcommand\ammonia{\ifmmode{{\rm NH}_{3}}\else{NH$_{3}$}\fi}
\newcommand\acetylene{\ifmmode{{\rm C}_{2}{\rm H}_{2}}
                        \else{C$_{2}$H$_{2}$}\fi}
                        
\begin{document}

\title{DREAM }
\subtitle{III. A helium survey in exoplanets on the edge  of the hot Neptune desert with GIANO-B at TNG}

\titlerunning{DREAM: III}

\authorrunning{Guilluy et al.}

\author{ G. Guilluy\inst{1},
         V. Bourrier\inst{2},
         Y. Jaziri\inst{2} , W. Dethier\inst{3}, D. Mounzer\inst{2}, P. Giacobbe\inst{1}, O. Attia\inst{2}, R. Allart\inst{4}\thanks{Trottier Postdoctoral Fellow}, 
         A.~S.~Bonomo\inst{1}, L. A. Dos Santos\inst{5}, M. Rainer\inst{6}, A. Sozzetti\inst{1}
}

\institute{
INAF -- Osservatorio Astrofisico di Torino, Via Osservatorio 20, 10025, Pino Torinese, Italy
\and
Observatoire Astronomique de l'Universit\'e de Gen\`eve, Chemin Pegasi 51b, 1290, Versoix, Switzerland
\and
Univ. Grenoble Alpes, CNRS, IPAG, 38000 Grenoble, France
\and
D\'epartement de Physique, Institut Trottier de Recherche sur les Exoplan\`etes, Universit\'e de Montr\'eal, Montr\'eal, Qu\'ebec, H3T 1J4, Canada
\and
Space Telescope Science Institute, 3700 San Martin Drive, Baltimore, MD 21218, USA
\and
INAF -- Osservatorio Astronomico di Brera, Via E. Bianchi, 46,
23807 Merate (LC), Italy
}

\date{Received date ; Accepted date }

\abstract
{The population of close-in exoplanets features a desert of hot Neptunes whose origin
remains uncertain. These planets may have lost their atmosphere, eroding into mini-Neptunes and rocky super-Earths below the desert. Direct observations of evaporating atmospheres are essential to derive mass-loss estimates and constrain this scenario. The metastable \ion{He}{I} triplet at 1083.3~nm represents a powerful diagnostic of atmospheric evaporation because it traces the hot gas in extended exoplanet atmospheres while being observed from the ground. In addition, it is located at the bright near-infrared stellar continuum and is very weakly affected by interstellar medium (ISM) absorption.}
{We carried out a homogeneous \ion{He}{I} transmission spectroscopy survey, targeting a selected sample of nine planets along the different edges of the desert, to interpret the absorption line profile with evaporation models and to better understand the role of photoevaporation in the desert formation.}
{We observed one transit per planet using the high-resolution, near-infrared spectrograph GIANO-B mounted on the Telescopio Nazionale Galileo telescope. We focused our analysis on the \ion{He}{I} triplet, based on a comparison of the in-transit and out-of-transit observations, and we computed high-resolution transmission spectra. We then employed the 1D  \textit{p-winds} model to calculate the planetary thermospheric structures and to interpret the observed transmission spectra.}
{We found no signatures of planetary absorption in the \ion{He}{I} triplet in any of the investigated targets. We thus provided 3~$\sigma$ upper-limit estimations on the thermosphere absorption, temperature and mass loss, and combined them with past measurements to search for correlations with parameters such as the stellar mass and XUV flux, which are thought to be key drivers in the formation of the \ion{He}{I} triplet.} 
{These results strengthen the importance of performing homogeneous surveys and analyses in bringing clarity to  \ion{He}{I} detections and (thereby) to plausible Neptunian desert origins. Our findings corroborate literature expectations that state the \ion{He}{I} absorption signal is correlated with the stellar mass and the received XUV flux. However, when translated in terms of mass-loss rates, these trends seem to disappear. Thus, further studies are essential to shed light on this aspect and to better understand  the photoevaporation process. }


\maketitle

\section{Introduction}

The population of close-in exoplanets ($P\approxinf$30\,days) features a dearth of Neptune-size planets on very short orbits ($P\approxinf$4\,days). This so-called ``Neptunian desert'' \citep[e.g.,][]{Lecavelier2007,Davis2009, Szabo2011,Beauge2013} is not an observational bias, as close-in Neptunes are easy to detect via both transits and radial-velocity measurements. Debate around the key driver mechanisms at the origin of the desert, which are linked to the formation, migration, and atmospheric evolution of close-in planets, is still ongoing. Photoevaporation \citep[e.g.,][]{Owen2018, Owen2019} and high-eccentricity orbital migration followed by tidal interaction with the star \citep[e.g.,][]{Matsakos2016} are the most likely explanations to date, but their interplay remains to be explored. Among questions that need to be addressed are the range of mass and period over which these processes are at play and whether they also shape the Neptunian ``savanna'' that is represented by a lighter deficit of Neptune-size planets at longer periods and lower irradiation, as highlighted by \citet{Bourrier2022}.

Investigating this complex puzzle is the goal of the Desert-Rim Exoplanets Atmosphere and Migration (DREAM) program. In DREAM I \citep{Bourrier2022}, we measured the orbital architectures of a large sample of exoplanets spanning the borders of the Neptunian desert and savanna. This work revealed a high fraction of misaligned orbits, strengthening the importance of high-eccentricity orbital migration for close-in planets. Architecture measurements from DREAM~I were included in a large statistical study of spin-orbit angles in DREAM II \citep{Attia2022}. This work confirmed the major role of tides in shaping the overall distribution of close-in planets' orbital architectures, except for a substantial fraction of planets on polar orbit that appears resilient to tidal realignment and further support the importance of disruptive dynamical processes. A subsample of the systems in DREAM I were observed in transit as part of a campaign our team led with the GIARPS observing mode (GIANO-B+HARPS-N) at the Telescopio Nazionale Galileo (TNG) telescope, to measure their Rossiter-McLaughlin effect in optical HARPS-N data and to analyze the planetary atmospheric spectra in near-infrared GIANO-B data. The objective of this third paper in the DREAM series is to search these GIANO-B data for absorption by helium escaping the upper atmosphere of these planets, bringing constraints on their mass loss and the role of atmospheric escape in the formation of the desert. 

Strong high-energy X-rays and extreme ultraviolet (XUV) stellar radiation can lead to an expansion of the upper atmospheric layers and the substantial escape of gas into space \citep[e.g.,][]{Vidal-Madjar2003, Lammer2003, Tripathi2015}. While hot Jupiters are generally stable against this photoevaporation, hot Neptunes have lower gravitational potential that makes them more vulnerable \citep[e.g.,][]{Lecavelier2004,Owen2017}. The upper atmospheric layers of these planets have traditionally been probed via transit spectroscopy in the ultraviolet (UV), by monitoring the change in absorption during transit of the stellar Ly$\alpha$ line. The \ion{H}{I} exospheres of hot Jupiters yield absorption signatures in the stellar Ly$\alpha$ that are ten times deeper than the lower atmosphere \citep{Lecavelier2012,Ehrenreich2012}, and this absorption level is even higher for the exospheres of warm Neptunes \citep[e.g.,][]{Ehrenreich2015,Lavie2017,Bourrier2018} -- and possibly even for mini-Neptunes \citep[e.g.,][]{dosSantos2020_, Zhang2022b}. However, UV observations can only be performed from space, and the stellar Ly-$\alpha$ line is contaminated by geocoronal emission and absorbed by the interstellar medium (ISM) absorption. While geocoronal emission can be reliably subtracted in the data reduction, there is nothing that can be done for the ISM absorption, so that only the wings of the line are usable when probing escaping atmospheres. In this way, gas dynamics in regions closer to the planet itself, at the wind-launching radius, remains obscured with Ly-$\alpha$ observations \citep[e.g.,][]{Murray-Clay2009,Owen2023} leading to low-precision mass-loss rates, as we observe the gas when it has already escaped in the exosphere. Additionally, Ly-$\alpha$ studies have only been performed on a few systems due to ISM absorption, which prevents observing stellar Ly-$\alpha$ lines beyond $\sim$50~pc. Accessing the thermosphere, the upper atmospheric layer below the exosphere, represents a way to overcome these limitations.

As it is very weakly affected by interstellar absorption and it can be observed from the ground the \ion{He}{I} triplet at $\lambda\sim$1083.3~nm (vacuum wavelength) has been recently identified as a robust alternative for tracing atmospheric expansion and evaporation \citep[e.g.,][] {Seager2000, Oklopcic2018}. The first detection of a helium thermosphere was obtained by \citealt{Spake2018} with the Wilde Field Camera 3 (WFC3) of the Hubble Space Telescope (HST). The \ion{He}{I} feature was not spectrally resolved due to the low-resolution of the data, however, further observations at high-resolution with CARMENES \citep[][]{Allart2019} allowed for the absorption lines  to be spectrally resolved and to derive atmospheric properties, showing that the helium tracer can probe the planet thermosphere and occasionally the exosphere. To date, \ion{He}{I} has been searched for in the
upper atmospheres of about 40 planets  which  (see Table~\ref{det_he}).

These observations have led to \ion{He}{I} absorption or upper limits estimations. However, the non-homogeneity in both observing methodology, and data reduction technique may mask possible trends in the data,  thus making it difficult to find a clear correlation with the parameters (e.g., stellar mass, XUV irradiation) that are believed to drive the detection (or non-detection) of the \ion{He}{I} triplet \citep[e.g.,][]{Fossati2022}.

Despite the several helium studies, the parameters that are considered to be important in triggering the \ion{He}{I} detection are currently under debate. For instance, \citealt{Oklopcic2018} proposed that planets orbiting K-type stars should be promising targets for showing evaporating or escaping helium atmospheres. This is because K-type stars emit a high amount of stellar XUV emission, which is responsible for \ion{He}{I} atoms ionization from the ground state (which can then recombine into the metastable state) and a low stellar mid-ultraviolet emission, which reduces the ionization of metastable \ion{He}{I} atoms \citep{Oklopcic2018}. However, the discovery of a strong helium signature for a gas-giant planet orbiting an F-star \citep{Czesla2022}, highlighted that also other planets orbiting stars with different spectral energy distribution (SED) can exhibit large helium outflow regardless of the stellar spectral type. More studies are thus essential to shed light on which are the mechanisms and parameters important in the \ion{He}{I} detection.

The sample we analyze in DREAM III is part of a pilot survey of nine planets located at the different edges of the Neptunian desert and savanna. The sample is described in Sect.~\ref{sample}, and its near-infrared (nIR) observations with the GIANO-B high-resolution spectrograph are presented in Sect.~\ref{observations}. We detail the data reduction procedures in Sect.~\ref{data_analysis}, and the interpretation of the helium absorption observations in Sect.~\ref{pwind}. We then present our findings in Sect. \ref{results}, followed by our conclusions in Sect.~\ref{summary}.

\begin{figure}
\centering
\includegraphics[width=0.78\linewidth,angle=270]{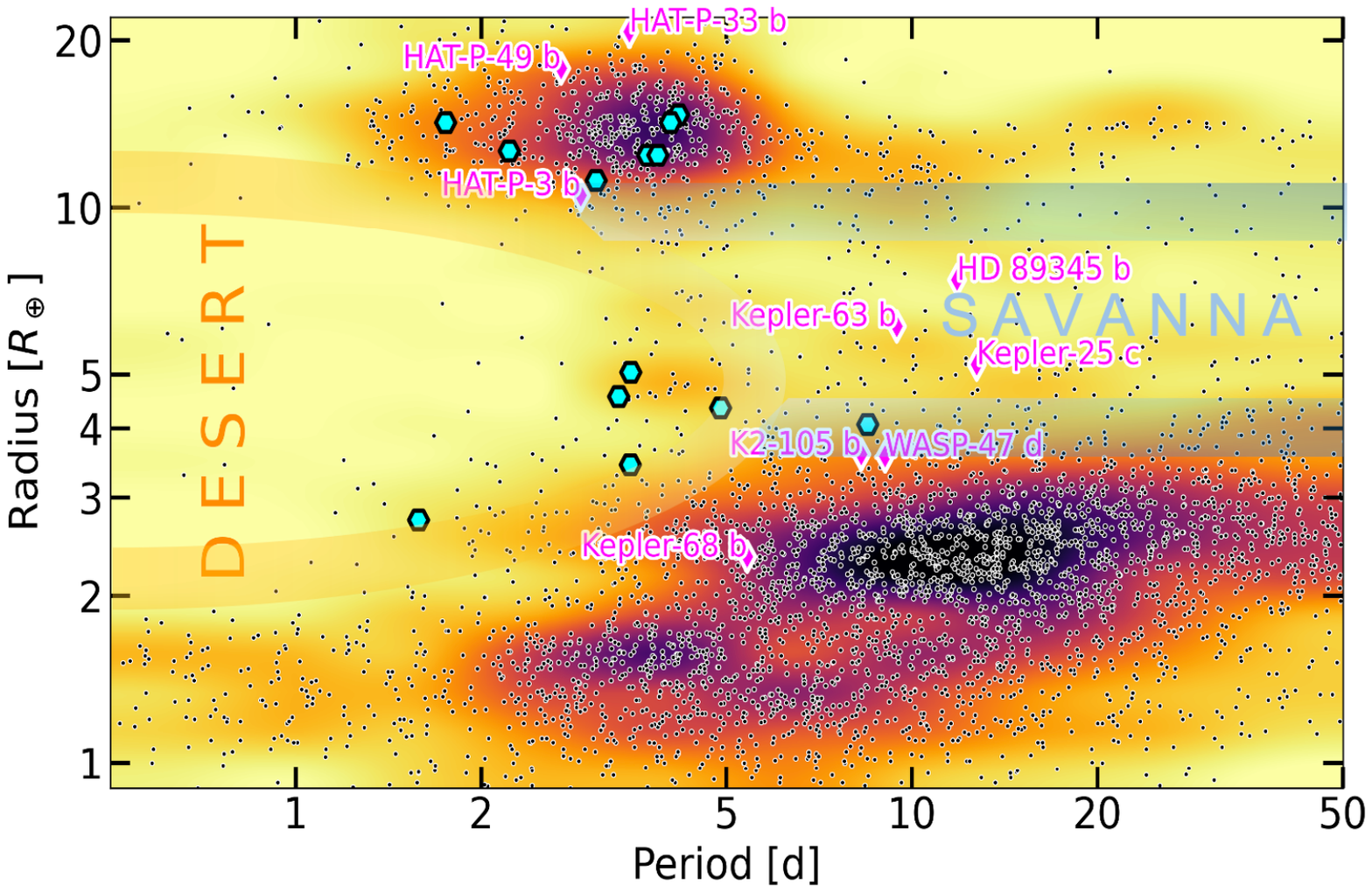}
\label{imm1}
\caption{Two-dimensional distribution of
exoplanets as a function of their
radius and period from the NASA  Exoplanet  Archive  \citep{Akeson2013}.
Magenta and pink diamonds represent DREAM III and \citet{Allart2023} -a work that we took as a reference for our analysis- targets, respectively. The approximate boundaries of the desert and savanna are highlighted. }
\end{figure}
\section{Sample} \label{sample}
Our survey consists of nine planets along the transitions defining the Neptunian desert and savanna. Their low densities ($\rho_\mathrm{pl}$<3.5 \gcm) and bright host stars (J<10.7) favor observations of their atmosphere in the nIR. The planet and star parameters adopted in this work are presented in Table~\ref{Tab_parameters}.  Below, we describe the main features of interest for these planets, which led to their inclusion in our survey. 
We would like to stress that no helium studies were reported in the literature for these investigated planets.\\
\textit{Hot Jupiter HAT-P-3b}. Its strong irradiation is expected to induce a large mass loss. The small radius of HAT-P-3b is indicative of a metal-enriched composition \citep{Chan2011}, which could be the result of atmospheric escape over the last 2.6 Gyr. Hydrogen would be lost preferentially, making helium a particularly interesting tracer for this planet.
HAT-P-3b orbits a K-star, thus according to \citet{Oklop2019}, it is likely to show metastable helium absorption.
DREAM I reports a polar orbit for this planet and, if confirmed, dynamical simulations will be needed to understand whether the present-day architecture is a result of a disruptive dynamical history (with partial evaporation of its volatile content) or a primordial misalignment between the protoplanetary disk and the star. \\
\textit{Hot Jupiter HAT-P-33b}. The extreme irradiation and very low density (0.134$_{-0.042}^{+0.053}$  g~cm$^{-3}$, \citealt{Wang2017}) of this highly inflated planet are expected to induce a large mass loss. \citet{Turner2017} measured an excess depth during transit in the R-band, which contains the H$\alpha$ transition, suggesting that the planet may be undergoing hydrodynamical escape. The misalignment of the system, due to the inclination of the host star, suggests that HAT-P-33 b underwent a high-eccentricity migration, and thus it possibly migrated close to the star long after its formation, which would change its atmospheric history compared to an early-on migration and erosion.\\
\textit{Ultra-hot-Jupiter HAT-P-49b}. It is a gas giant exoplanet discovered orbiting a bright (V = 10.3) slightly evolved F-star \citep{Bieryla2014}. Its extreme irradiation, due to its proximity to the host star (a=0.0438$\pm$0.0005~au, \citealt{Bieryla2014}), is expected to induce a large mass loss. According to the analysis presented in DREAM I, the planet is probably on a polar orbit, supporting a disruptive dynamical origin or evolution for the system, whose architecture was unaffected by tidal interactions with the shallow convective envelope of the host star (DREAM II).\\
\textit{Warm super-Neptune HD89345b}. This planet is five times more irradiated than the evaporating super-Neptune WASP-107b, yet it survived atmospheric escape for 9.4~Gyr \citep[e.g.,][]{Van2018}. HD~89345b stands at the transition between stable
Jupiter-mass planets and hot Neptunes that entirely lost their atmosphere and this is thus an essential piece in the puzzle that is the origin of the desert. HD~89345b is located on a misaligned orbit (DREAM I) right  within  the  savanna (see Fig.~\ref{imm1}). The present-day misalignment could trace both a primordial formation of the system, arising from the tilt of the early star or protoplanetary disk, or the planet could have migrated more recently, exiting a Kozai resonance with an outer companion \citep[e.g.,][]{Bourrier2018, Attia2021}. This second scenario would imply that HD~89345b arrived near the star at the end of its main-sequence lifetime, changing our view of its irradiative history and our interpretation of its inflation \citep{Yu2018} and hydrodynamical escape.\\
\textit{Warm sub-Neptune K2-105b}. It remains unclear why sub-Neptunes appear to be more resilient than warm Neptunes to the processes that created the desert \citep{Owen2019}. K2-105b stands at the transition between these two populations and is predicted to have an atmosphere accounting for up to 10\% of its total mass \citep{Narita2017}.
Detecting the presence of this atmosphere and measuring its mass loss could bring constraints on the interior of the planet; if its evolution was controlled by atmospheric escape, it is estimated to have retained its envelope only if its core mass is greater than 6~\mearth \citep{Narita2017}. 
DREAM I reported a possibly misaligned orbit which, if confirmed, might support a turbulent dynamical history and the planet's late arrival into its close-in orbit. However, the presence of other targets may indicate a primordial inclination of the star or protoplanetary disk, as K2-105b is far away from its host stellar companion to  experience tidal interactions.\\
\textit{Warm Neptune Kepler-25c}. It is close to a resonant periodic configuration with a companion planet, which is known to be the final state of a system that undergoes migration within the protoplanetary disk \citep{Migaszewski2018}. Kepler-25c should thus be evaporating since its formation 11 Gyr ago \citep{marcy2014}, yet its low density (0.588$^{+0.053}_{-0.061}$~g~cm$^{-3}$, \citealt{Mills2019}) indicates the presence of a H/He envelope.\\
 \textit{Warm Neptune Kepler-63b}. It is a gas giant exoplanet with a radius between Neptune and Saturn.  The orbital period is around 9.4 days, leading to an equilibrium temperature of about 900 K \citep{Mallonn2022}.  The planet is in a polar orbit around a young Sun-like star (\citealt{Sanchis-Ojeda2013}, DREAM I), thus offering the possibility to assess how evaporation shapes a Neptune’s atmosphere in its early life. Its radius and insolation are similar to those of the other Neptunian targets, but it is much younger (200 Myr vs 10 Gyr) and still possibly undergoing vigorous escape. \\
 \textit{Hot sub-Neptune Kepler-68b}. With a density of 3.32$^{+0.86}_{-0.98}$ \gcm \citep{Gilli2013}, it is considered a candidate ocean planet \citep{Zeng2014}
possibly topped by a moderate H/He envelope \citep{Howe2014}. Detecting He would offer insights on the mysterious
nature of this sub-Neptune, representative of the transition between rocky planets and gas giants \citep{Lopez2014}. \\
\textit{Warm sub-Neptune WASP-47d}. The WASP-47 planetary system is composed of at least four planets, a hot Jupiter (WASP-47 b; P = 4.159 days, \citealt{Bryant2022}) with an inner super-Earth (WASP-47 e; P = 0.7896 days, \citealt{Bryant2022}), a close-orbiting outer Neptune (WASP-47 d; P = 9.031 days, \citealt{Bryant2022}), and a long-period giant planet (WASP-47 c; P = 588.4 days, \citealt{Vanderburg2017,Bryant2022}). WASP-47 d is near a 2:1 resonance with the inner Hot Jupiter WASP-47b.
It has a similar radius and insolation of K2-105b but is three times less massive. Their comparison could provide valuable insight into evaporation processes on sub-Neptunes.

\section{Observations} \label{observations}
\begin{figure}
\centering
\includegraphics[width=\linewidth,height=18cm]{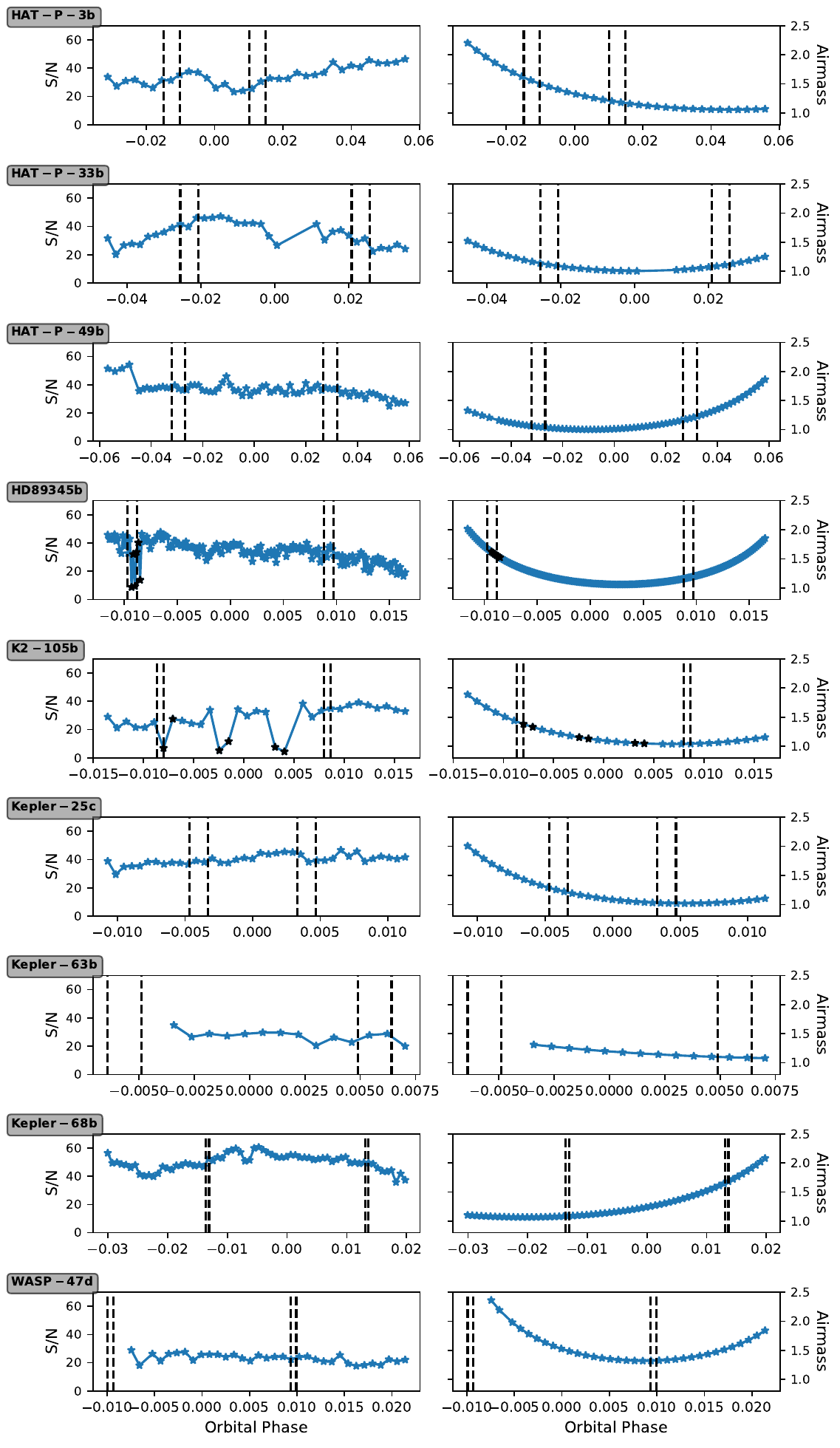}
\caption{S/N in the region of interest (1082.2-1085.5~nm)(left panel) and airmass (right panel) measured during the GIARPS observations. The vertical dashed lines mark the $t_1$, $t_2$, $t_3$, and $t_4$ contact points (from left to right). Black stars indicate the discarded AB couples for low S/N. }
\label{snr_am_}
\end{figure}
\begin{table}
        \caption{Log of TNG-GIANO-B observations.}
        \label{tab_log} 
        \centering  
        \resizebox{\linewidth}{!}{
        
        \begin{tabular}{l | c | c | c | c | c }          
                \hline\hline                       
                  Target & Night & S/N$_{\rm{AVE}}$ & am$_{\rm{min}}$-am$_{\rm{max}}$ & N$_{\rm obs}$ & Exposure time [s] \\  
                \hline  
 HAT-P-3b & 2020-01-30 & 34 & 1.1-2.2 & 34 & 600.0 \\
 HAT-P-33b & 2019-12-04 & 35 & 1.0-1.5 & 34 & 600.0 \\
 HAT-P-49b & 2020-07-30 & 37 & 1.0-1.9 & 72 & 300.0$^+$ \\
 HD89345b & 2020-02-02 & 34 & 1.1-2.0 & 176 & 100.0 \\
K2-105b & 2020-01-18 & 27 & 1.0-1.9 & 32 & 600.0 \\
Kepler-25c & 2019-06-14 & 40 & 1.0-2.0 & 38 & 600.0 \\
Kepler-63b & 2020-05-13 & 27 & 1.1-1.3 & 14 & 600.0 \\
Kepler-68b & 2019-08-03 & 50 & 1.1-2.1 & 66 & 300.0 \\
WASP-47d & 2021-07-30 & 23 & 1.3-2.4 & 34 & 600.0 \\
                \hline 
        \end{tabular}
        }
        \tablefoot{
                 time-averaged S/N in the spectral region containing the He{\sc i} triplet (1082.2--1085.5\,nm). \\($^+$) First four exposures at 600s.}
\end{table}
We observed the systems in our sample with the nIR echelle spectrograph GIANO-B installed on the 3.6~m Telescopio Nazionale Galileo (TNG) telescope. The observations were performed with the GIARPS  configuration and were carried out with the nodding acquisition ABAB \citep{GIARPS_claudi}. Therefore, while the targets were observed in one nodding position along the slit (A and B), the sky spectra were gathered simultaneously with the other one, thus providing an accurate reference for subtracting the thermal background and telluric emission lines.

GIANO-B covers the Y, J, H, and K spectral bands (0.95-2.45~$\mu$m) in 50 orders at a resolving power of R$\sim$50,000. For this analysis, we focus on order \#39, where the helium triplet falls. We collected a transit observation for each investigated target. The only exception is HAT-P-3b, as due to bad weather conditions, we collected two nights of observation, namely UT 14 April 2019 and UT 30 January 2020, but the first visit was
excluded from our analysis since observations had to be stopped
just before the transit. A log of the observations is reported in Table~\ref{tab_log}. Figure~\ref{snr_am_} shows the variation in the signal-to-noise ratio (S/N) for  order \#39 and the variation in airmass for each exposure. Due to the lack of a sufficient number of collected images, we had to discard Kepler-63b from our analysis. Moreover, K2-105b's and HD89345b's observations were affected by GIANO-B auto-guide problems and by the presence of clouds, 
so we decided to discard the AB couples of exposures
which exhibit a very low S/N.

\section{Data analysis} \label{data_analysis}
Extended or evaporating atmospheres can be detected through an excess absorption by metastable helium in the planet transmission spectrum. In the following section, we discuss the steps we performed in order to reduce the raw GIANO-B data, extract individual transmission spectra, and calculate average in-transit spectra in the planet rest frame.\\
\subsection{Initial data reduction} \label{initial_red}
The raw spectra were dark-subtracted, flat-corrected, and extracted (without applying the blaze function correction) using the GOFIO pipeline \citep{Rainer2018}. In addition, GOFIO yields a preliminary wavelength calibration (defined in vacuum) using U-Ne lamp spectra as a template. We used the ms1d spectra, with the echelle orders separated and the Barycentric Earth Radial Velocity (BERV) correction applied, the spectra are defined in the terrestrial rest frame.

Since the U-Ne is acquired at the end of the night to avoid the persistence of the saturated signal of some emission lines on the detector polluting the scientific observations, the mechanical instability of GIANO-B makes the wavelength solution determined by GOFIO insufficient in terms of accuracy. We corrected for this by aligning all the GIANO-B spectra to the telluric reference frame via spline-interpolation based on the retrieved shifts obtained by cross-correlating with a time-averaged spectrum used as a template \citep{Brogi_2018_Giano,Guilluy2019,Guilluy2020, Giacobbe2021}. We thus aligned the spectra to the reference frame of the Earth’s atmosphere, which is also assumed as the frame of the observer (neglecting any $\sim$10 m s$^{-1}$ differences due to winds). We then used the atmospheric transmission spectrum generated via the ESO Sky Model Calculator\footnote{\url{https://www.eso.org/observing/etc/bin/gen/form?INS.MODE=swspectr+INS.NAME=SKYCALC}} to refine the initial GOFIO wavelength calibration.\footnote{When the telluric lines are not strong enough, the re-alignment into the telluric rest frame may not work properly, as in the case of Kepler-68b. Thus, we preferred to discard this step in the analysis for this specific target.}
\subsection{Transmission spectroscopy}
We performed the transmission spectroscopy, applying the steps described below to each transit and target independently and considering the system parameters listed in Table~\ref{Tab_parameters}.

\subsubsection{Telluric correction}
First, we performed a detailed correction for telluric contamination.  
We used the {\fontfamily{pcr}\selectfont Molecfit} ESO software \citep{Smette2015,Kausch2015} to correct for the transmission telluric lines \citep{Allart2017}. As this is the first time that {\fontfamily{pcr}\selectfont Molecfit} has been applied to GIANO-B data, we report the adopted parameters in Table~\ref{tab_molec} .

 \begin{figure}

\includegraphics[width=\linewidth]{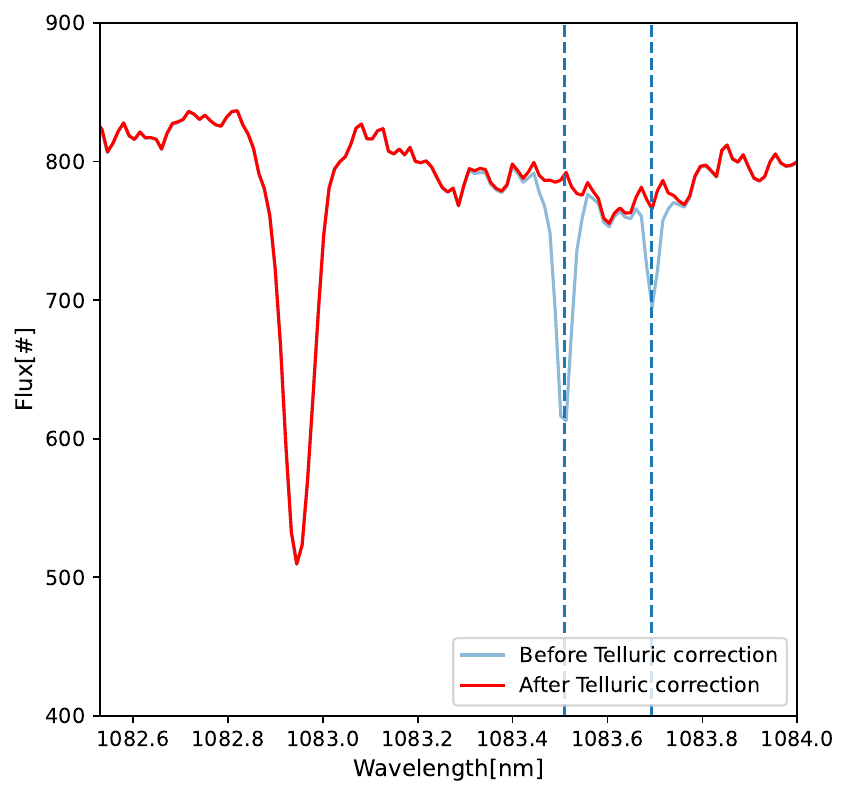}
\caption{Example of a time-averaged spectrum before (in blue) and after the telluric lines removal (in red). Dashed vertical lines highlight the position of two H$_2$O telluric lines, i.e., 1083.51~nm and 1083.69~nm, in the region around the \ion{He}{I} triplet. These two lines are corrected in the red spectrum. }
\label{star_frame}
\end{figure}
{\fontfamily{pcr}\selectfont Molecfit} is based on a combination of two different sources: an atmospheric standard profile (MIPAS),  and a Global Data Assimilation System (GDAS) profile. {\fontfamily{pcr}\selectfont Molecfit} gives the merging of these two profiles as input for a line-by-line radiative transfer model (LBLRTM). We considered the precipitable water vapor in our transmission model, and selected a fixed grid to merge the two atmospheric models, namely, the variations in temperature, pressure, humidity, and abundance of H$_2$O from 0 to 120 km are described with a fixed number of layers (50). Meanwhile, LBLRTM  returns the telluric spectrum. We considered one observation at a time,and we initially performed the model fitting on selected spectral intervals inside the order \#39 showing a well-determined continuum level, a good number of telluric lines, and a few or zero stellar lines. Based on the best-fit parameters derived by {\fontfamily{pcr}\selectfont Molecfit,} we then generated a telluric spectrum for the entire spectral order and we corrected the science spectrum. An example of telluric removal is shown in Fig.~\ref{star_frame}.

In the spectral region of interest, there are three OH emission lines that fall near the \ion{He}{I} triplet (at $\sim$1083.21~nm, $\sim$1083.24~nm, and $\sim$1083.43~nm, vacuum wavelengths). As the observations were gathered with the nodding acquisition mode that allows for the subtraction of the thermal background and emission lines (see Sect.~\ref{initial_red}), there is no need to correct for telluric emission lines. However, due to seeing variations during the observing nights, the A-B subtraction can leave some residuals at the wavelengths of the OH lines. We thus masked the correspondent wavelengths. 

\subsubsection{Alignment into stellar rest frame.} We then shifted the  spectra in the stellar rest frame by accounting for the stellar radial velocity, $V_{\star}$, in the telluric reference system. This is given by:
\begin{equation}
V_{\star/\oplus}=\sum_{i}{ K_{\star i}[cos (\nu_i + \omega_i)} + e_i cos(\omega_i)]+V_\mathrm{sys} + V_\mathrm{bar},
\end{equation}
 where we account for the velocity of the observer induced by the rotation of the Earth and by the motion of the Earth around the Sun, namely:\ the barycentric Earth radial velocity, $V_\mathrm{bar}$, the stellar reflex motion induced on the host star by each planet $i$ in the system (i.e. $K_{\star i}[cos (\nu_\mathrm{i} + \omega_\mathrm{i}) + e_\mathrm{i} cos(\omega_\mathrm{i})]$, where $\nu_\mathrm{i}$ is the true anomaly obtained from the eccentric anomaly via the Kepler’s equation, in this way we directly account for the eccentricity which is significative for some of our targets, e.g., HAT-P-33b, HD89345b),  $\omega_\mathrm{i}$ is the argument of periastron, $e_\mathrm{i}$ is the eccentricity, and $K_{\star i}$ is the stellar radial-velocity semi-amplitude, and the systemic velocity of the star-planets system with respect to the barycentre of the Solar System ($V_\mathrm{sys}$). 
\begin{figure*}
\centering
\includegraphics[width=0.95\linewidth]{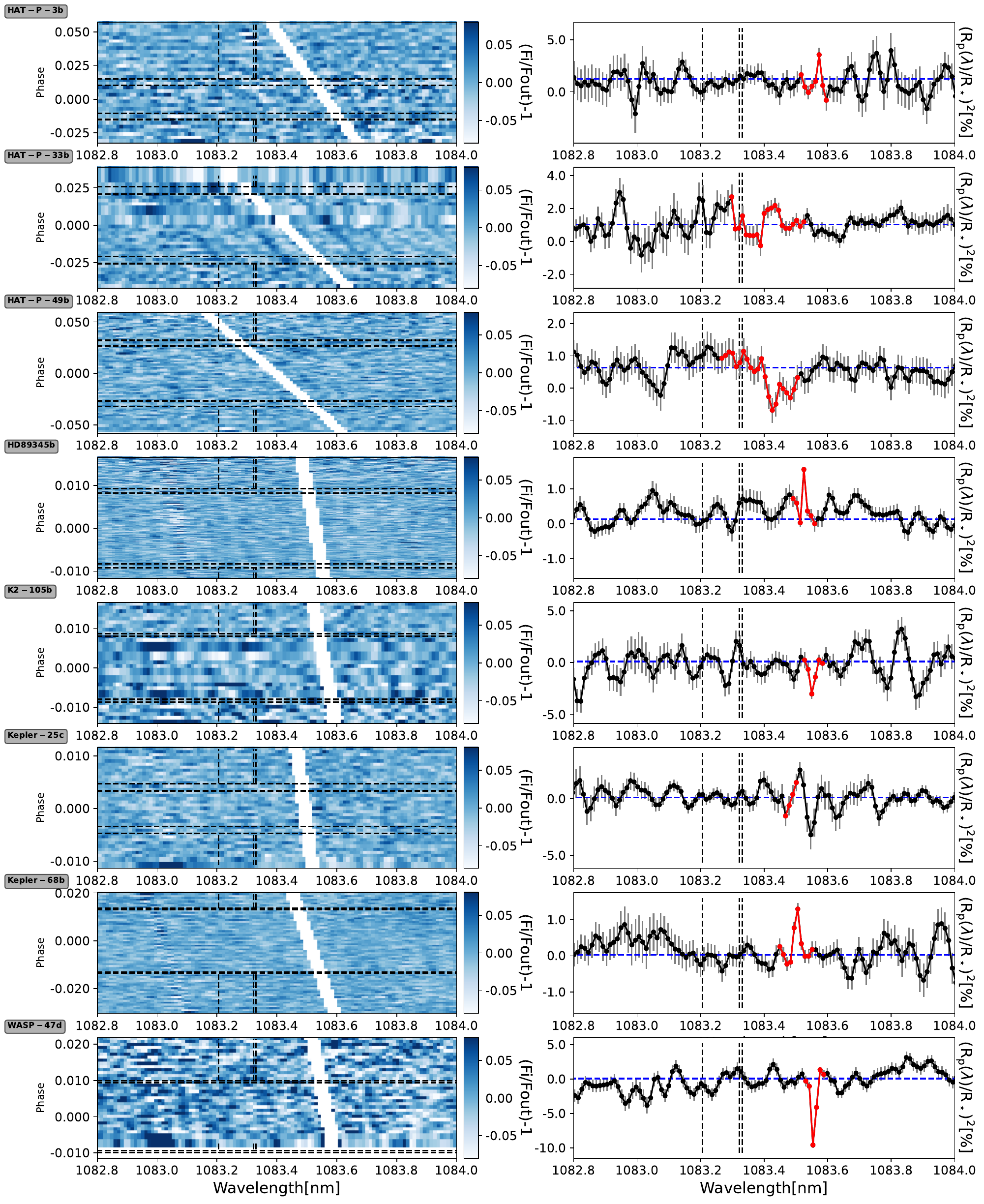}
\caption{ Transmission spectra (T$_\mathrm{\lambda,i}$=$\tilde{\mathrm{F}_\mathrm{i}(\lambda)}$/S$_\mathrm{out}(\lambda)$ -1) shown in tomography in the planetary rest frame in the region of the \ion{He}{I} triplet, as a function of wavelength and planetary orbital phase (left). The contact points t$_1$, t$_2$, t$_3$, and t$_4$ are marked with horizontal black lines. The regions affected by OH$^{-}$ contamination are masked.  Mean-transmission spectrum for
each observed transit (right). The horizontal blue line is the white-light radius $(R_\mathrm{pl}/R_\star)^2$. The mean transmission spectra have an inverted sign compared to $\tilde{\mathrm{F}_\mathrm{i}(\lambda)}$/S$_\mathrm{out}(\lambda)$ -1 as the radius is expressed instead of absorption.  Black vertical lines indicate the position of the \ion{He}{I} lines. Red line marks the spectral regions affected by OH$^{-}$ emission. For some planets, some residuals are left at the position of the Si line ($\sim$1083~nm). This is due to the depth of the line which can give rise to difficulties in the spectral extraction (see e.g.,  \citealt{Krishnamurthy2023})}.
\label{tomo}
\end{figure*}

\subsubsection{Transmission spectra calculation.} For every considered target, we divided each spectrum\footnote{We did not consider some spectra which exhibited much lower S/N compared to the other exposures or outliers near the position of the \ion{He}{I} triplet -see Fig.\ref{App_bad} for details.} by its median value, thus obtaining the normalized spectra, $\tilde{\mathrm{F}_\mathrm{i}}$. We then built a master stellar spectrum, $S_\mathrm{out}(\lambda)$, by averaging the out-of-transit spectra (i.e., with an orbital phase smaller than t$_1$ or greater than t$_4$), and derived individual transmission spectra, $T_\mathrm{\lambda,i}$, by dividing each spectrum by $S_\mathrm{out}(\lambda)$, that is, T$_\mathrm{\lambda,i}$=$\tilde{\mathrm{F}_\mathrm{i}(\lambda)}$/S$_\mathrm{out}(\lambda)$ -1.
Finally, we linearly interpolated transmission spectra in the planet rest frame, as follows:
\begin{equation}
\begin{split}
V_{\mathrm{pl_j/\star}}& =-\sum_{i}{ K_{\star i}[cos (\nu_i + \omega_i)+ e_{i} cos(\omega_i)]}\\
& -K_{\mathrm{pl_j}}(cos(\nu_j+\omega_j)+e_j cos(\omega_j) )    ,
\end{split}
\end{equation}
where $K_{\mathrm{pl_j}}$ is the computed planet radial-velocity semi-amplitude (see Table~\ref{Tab_parameters}) of the considered target $j$). 
The 2D maps of the transmission spectra in the planet rest frame are shown in the left panels of Fig.~\ref{tomo}.\\

The usual method to search for faint planetary atmospheric signatures is to average in-transit transmission spectra in the planet rest frame and thus boost the S/N. However, the naive calculation of transmission spectra performed above neglects the change in broadband flux level of in-transit flux spectra, due to the occultation of regions with varying flux intensity by the opaque planetary disk. For example, limb-darkening, if unaccounted for, biases the retrieved atmospheric absorption signal toward smaller values at phases close to the stellar limb, as compared to the stellar disk center. We thus followed the approach presented in \citet{Dany2022} that had been re-adapted from \citet{Wyttenbach2020} to compute the in-transit transmission spectra:
\begin{equation}
\label{eq_T}
    \left(\frac{R_\mathrm{pl}(\lambda,t)}{R_\star}\right)^2=\frac{LD_\mathrm{mean}}{LD(t)}\frac{F_\mathrm{out}(\lambda)-(1-\delta(t))F_\mathrm{i}(\lambda,t)}{F_\mathrm{local}(\lambda,t)}
,\end{equation}
where $F_\mathrm{i}(\lambda,t)$ is each observed spectrum at phase $t$, $1-\delta(t)$ is the broadband ("white-light") transit light curve, $\delta(t)$ is  the  transit  depth,
$LD(t)$ and $LD_\mathrm{mean}$ represent the stellar limb darkening at the position of the transiting planet and the disk-averaged limb-darkening, respectively\footnote{The limb-darkening correction is applied on the spectra aligned in the planet rest frame}. We used the Python $batman$ code  \citep{batman}  and  the  system  parameters  from  Table~\ref{Tab_parameters} to calculate the white-light transit light curve and the limb-darkening coefficients (see Fig.~\ref{LC_panels}). $F_\mathrm{local}(\lambda)$ is the normalized local  stellar  spectrum (see Sect.~\ref{ros_sect})  occulted  by  the  planet  at  phase $t$.
We applied Eq.~\ref{eq_T} only to fully in-transit orbital phases (i.e.,
obtained between the t$_2$ and t$_3$ contact points), while ingress, and egress were not considered here. Indeed LD, and occulted stellar surface are not well known at the limbs. If we neglect the RM effect and center-to-limb (CLV) variations, the occulted local stellar spectrum $F_\mathrm{local}(\lambda)$ is equal to the disk-integrated stellar spectrum $F_\mathrm{out}(\lambda)$ (see Sect.~\ref{ros_sect}), so that Eq.~\ref{eq_T} becomes:
\begin{equation}
\label{eq_T2}
\left(\frac{R_\mathrm{pl}(\lambda,t)}{R_\star}\right)^2=\frac{LD_\mathrm{mean}}{LD(t)}\left(1-\frac{(1-\delta(t))F_\mathrm{i}(\lambda,t)}{F_\mathrm{out}(\lambda,t)}\right).
\end{equation}

All transmission spectra fully in-transit were finally
averaged (T$_\mathrm{mean}$) to create one transmission spectrum for each observed transit (right panels of Fig.~\ref{tomo}).

\subsubsection{Fringing correction}
 Our GIANO-B spectra presented a sinusoidal fringing pattern caused by the sapphire substrate ($\sim$0.38~mm thick) placed above the sensitive part of the detector, which behaves as a Fabry-Pérot in generating interference fringing. Such fringing patterns must be corrected for in studying the \ion{He}{I} triplet. We followed and re-adapted the second and third approaches (Method\#1b-Method\#2) presented in \citet{Guilluy2019}. We focused on correcting this effect at the level of the final transmission spectra in order to have better control over fringing in the final transmission spectrum itself and to avoid risks in "overfitting" the data. First, for each planet, we binned T$_\mathrm{mean}$ (bin size of 0.2~nm), we thus computed the Lomb-Scargle periodogram to find the characteristic frequency of the periodic fringing signal present in the data $f_\mathrm{best}$. We then selected the most prominent frequency of the periodogram, and we fitted the fringing pattern using a sine function $yfit=C+A\sin(2 \pi \lambda f + \phi)$, where $A$ is the amplitude, $\phi$ the phase, $f$ is the fringing frequency (where we assumed $f_\mathrm{best}$ as starting point for the fit), and $C$ is the overall offset. We finally corrected our final transmission spectra by $yfit$.

\begin{figure*}
    \centering
    \includegraphics[width=\linewidth,height=18cm]{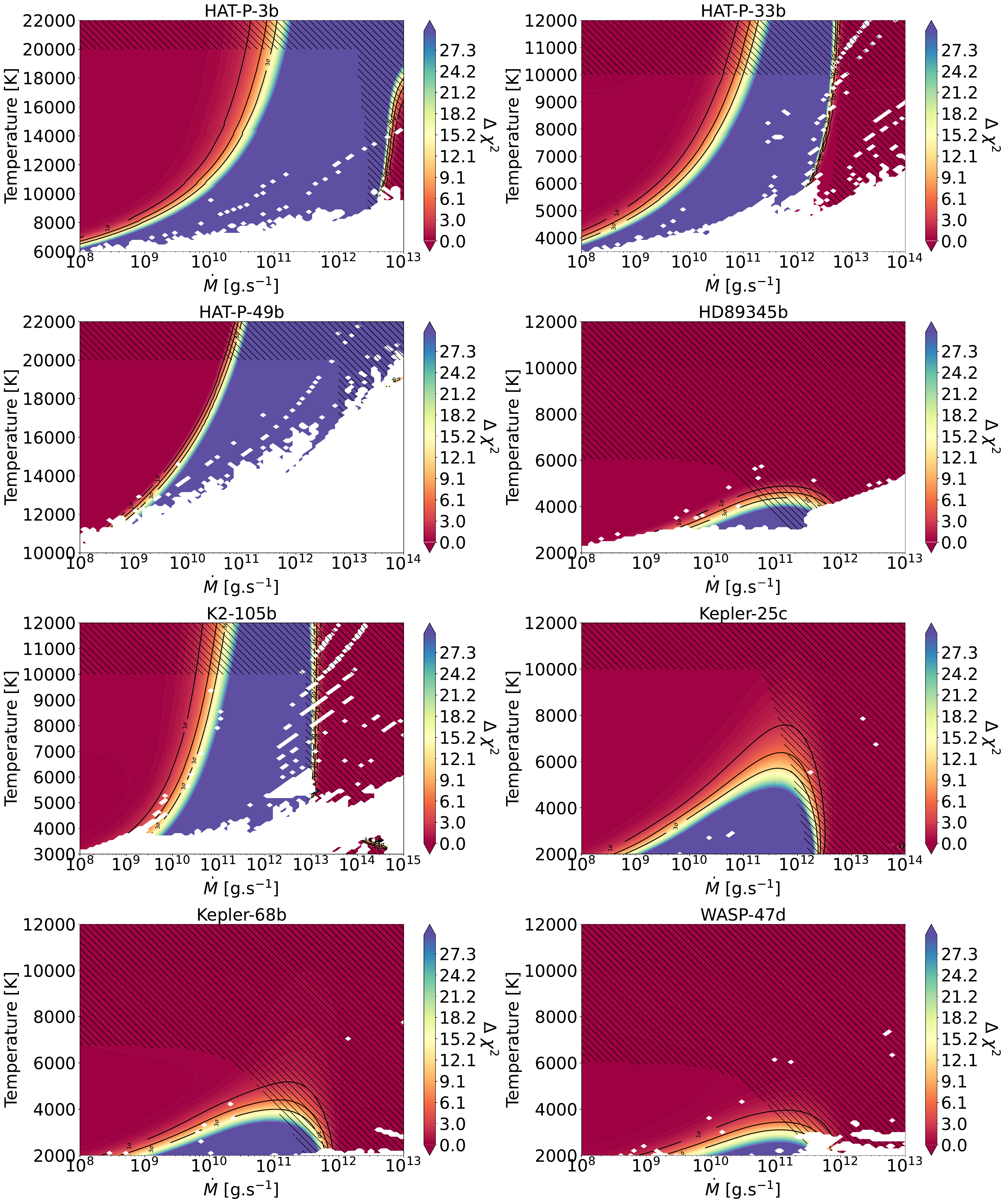}
    \caption{$\Delta\chi^2$ maps of mass-loss rate and temperature for the planets of our survey. Regions of the parameter space in red are consistent with the non-detections in our data, while models in purple are strongly disfavored. The thermospheric model failed to converge in white regions. Hatched regions are physically excluded. }
    \label{chimap}
\end{figure*}
\section{Interpretation of the transmission spectra} \label{pwind}

Somewhat surprisingly no \ion{He}{I} absorption signature was detected in our sample, as can be seen in Fig.~\ref{tomo}. This either means that the targeted planets have no extended atmosphere, which would be surprising given the strong irradiation of their H/He atmosphere; or that their thermosphere's metastable helium population is not dense enough to be detectable within the precision of the GIANO-B observations. Under this assumption, we can still put upper limits on the escape rate by fitting the transmission spectra with models of the planets' thermospheric structure.

\subsection{Stellar modelling}
\label{stelmod}

Given the scarcity of stellar high-energy measurements, we calculated the X-EUV spectral energy distribution of the eight target stars in a consistent manner using Table 5 in \citealt{linsky2014}. This formula depends on the total X-EUV flux emitted by the star, which is calculated based on the stellar age and Equations 3 and 4 from \citealt{sanz-forcada2011}.

\subsection{Thermosphere modelling}

We used an approximate 1D model, the \textit{p-winds} \citep{dossantos2022} code, largely based on the formulations of \cite{Oklopcic2018} and \cite{lampon2020}, to calculate the thermospheric structure and resulting signature of the metastable helium triplet. The atmospheric density and velocity profiles were calculated according to the Parker wind approximation, assuming an isothermal planetary outflow \citep{parker1958}. We assumed for all targets an atmospheric composition of 90~\% H and 10~\% He (a good approximation of the Jupiter H/He ratio) and an input stellar X-EUV spectrum  (calculated as explained in Section \ref{stelmod}). The code calculates only the density profiles of hydrogen in its neutral and ionized states, as well as that of helium in its neutral, excited, and singly ionized states. The signature of interest is the metastable transition at 1083.3~nm of the helium excited level. A theoretical ideal spectrum is calculated at mid-transit without taking into account geometrical effects and inhomogeneities of the stellar surface. This absorption signature is compared to the observed mean transmission spectrum to estimate upper atmosphere characteristics such as temperature and the mass-loss rates.

Since there is no clear evidence of a helium signature, we quickly explored the input parameter space of the \textit{p-winds} models by varying only the isothermal temperature profile, $T$, and the total atmospheric escape rate, $\dot{M}$, while all other input parameters are fixed. For instance, H/He ratio was not a fitting parameter. We expect that this can slightly change the derived values, but the conclusion would still remain the same. In our models, the radius at the top of the simulated atmosphere was set to the Roche lobe \citep{Eggleton1983}. We note that the value chosen for this upper radius has not been discussed in previous studies using the \textit{p-winds} code or similar codes \citep[e.g.,][]{Oklopcic2018,lampon2020,dossantos2022,Kirk2022}, even though it directly controls the amount of helium that contributes to the theoretical absorption signature. The preferred approach in the literature seems to be increasing the radius until the neutral triplet helium density no longer contributes significantly to the absorption signal. Yet this is hardly compatible with the change in nature of the atmosphere beyond the Roche lobe, from a collisional thermosphere shaped by planetary gravity, which may (at first order) still be described by a 1D vertical structure, to an asymmetrical exosphere shaped by the stellar gravity, radiation, and wind. Our choice to set the upper model radius at the Roche lobe is based on the reasonable assumption that once helium atoms escape into the exosphere, they cannot be excited into their metastable state by collisions anymore and are quickly photo-ionized so that these layers contribute little to the observed signature (as supported by the lack of clear detection of extended exospheric tails in the literature). 

In our simulations, high escape rates lead to an increase in the total density of metastable helium in the thermosphere but the densest layers are shifted to higher altitudes above the Roche lobe, where they no longer contribute to the theoretical signature. This boundary effect is visible in Fig.~\ref{chimap}, with the reappearance of fit regions compatible with the non-detection of an absorption signature in our data. Simulations at high escape rates are therefore model-biased and should be considered cautiously. Furthermore, we note that the code \textit{p-winds} was unable to calculate the atmospheric structure in certain regions of the parameter space (shown in white in Fig.~\ref{chimap}). It is still unclear whether this is a numerical issue or a truly non-physical regime for the thermosphere.

\subsection{Parameter space exploration}

We determined whether the models were compatible with the measured transmission spectra using $\chi^2$ comparison. Since no absorption signature was detected for any of the planets we took the null hypothesis (a flat transmission spectrum) as the best-fit model and use $\Delta\chi^2 = \chi^2_{model} - \chi^2_{flat}$ as a criterion to determine 3-$\sigma$ upper limits on the atmospheric mass-loss rate. We constrained the parameter space to realistic models in mass loss, using the maximum efficiency for a photoionization-driven isothermal Parker wind \citep{Vissapragada2022}, and in temperature, using the model of \cite{salz2016} as a function of the gravitational potential of the planet. Below log($-\Phi_G$) = log $GM_{pl}/R_{pl}$ = 13.0~erg$\cdot$g$^{-1}$, their model predicts temperatures lower than 10\,000\,K, while above this limit, it predicts temperatures lower than 20\,000\,K. $\chi^2$ maps as a function of mass loss and temperature are shown in Fig.~\ref{chimap} for all planets in our sample. Table~\ref{table_result} gathers all the derived 3-$\sigma$ upper limits.

\section{Accurate stellar line profiles} \label{ros_sect}
\begin{figure}
\includegraphics[ width =\linewidth]{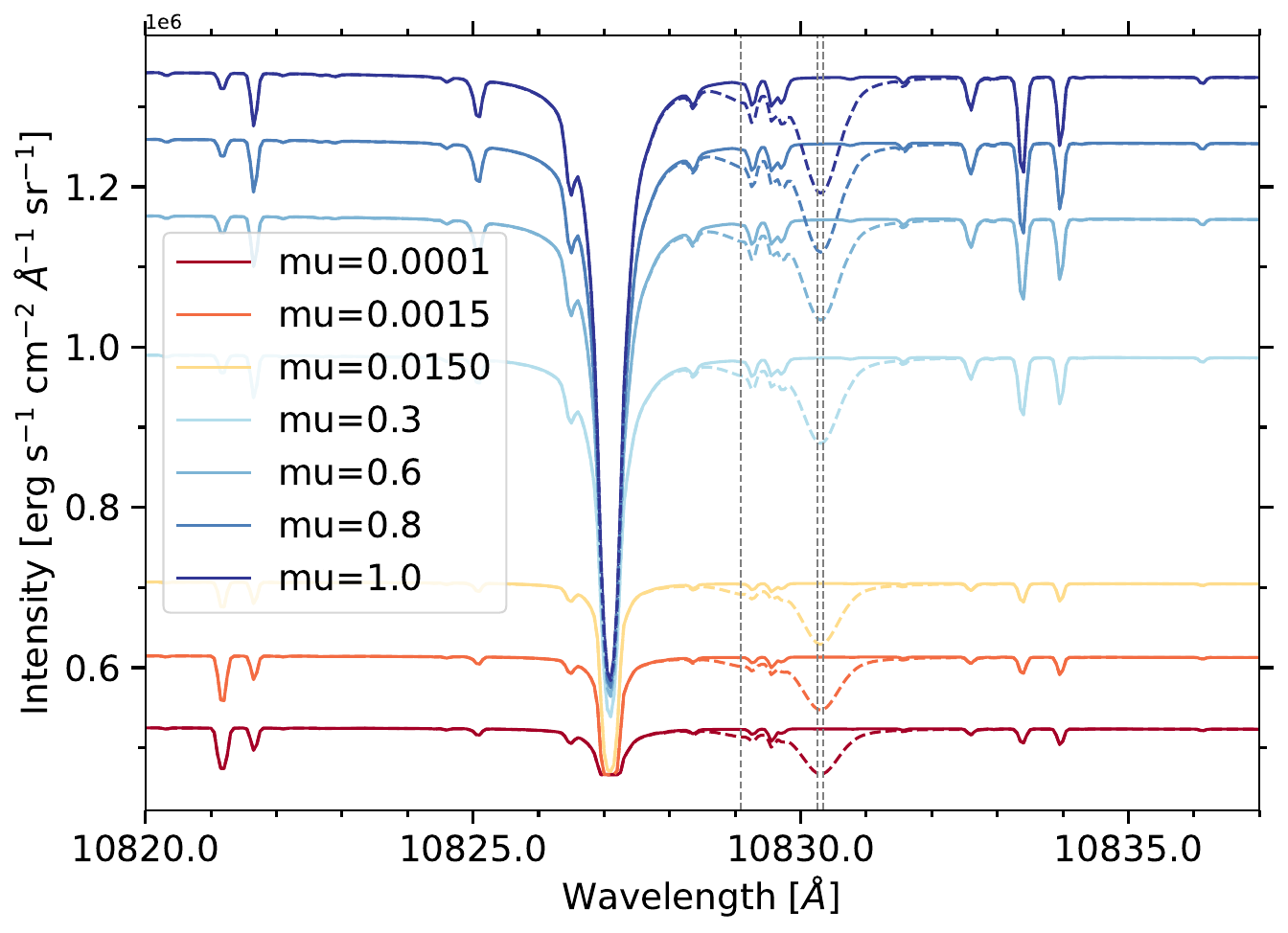}
\caption{Spectral intensities for different $\mu$ positions on the stellar disk of HAT-P-33 derived with \texttt{Turbospectrum}. The dotted spectra are obtained after the multiplication of the intensity spectra by the He\,I triplet spectral profile. $\mu = \sqrt{1 -(x^2 + y^2)}$ with $(x, y)$ the coordinates of a point on the stellar disk of radius equal to one in the Cartesian referential centered on the stellar disk.}
\label{fig:HAT_P_33_Imu}
\end{figure}
Planet-occulted line distortions (POLD, \citealt{Dethier2023}) can bias or even hide planetary absorption signatures in transmission spectra \citep{yan2017,casasayasbarris2020,casasayasbarris2021}. They appear in particular when one uses the disk-integrated stellar spectrum (F$_\mathrm{out}$) to normalize the spectrum that is absorbed by the planet and its atmosphere. Indeed, the line profiles of F$_\mathrm{out}$ are shaped by a combination of the local effects of stellar rotation and CLV from all over the stellar disk; thus, they are  not necessarily representative of the line profiles occulted by the planet. To mitigate the POLDs one needs to define more accurate estimates of the local stellar spectrum occulted by the planetary disk at each exposure. However, this quantity is complex to estimate from observations, as stars cannot be resolved spatially. \\

To estimate the planet-occulted stellar spectrum, we fit a model to the measured disk-integrated spectrum, using a combination of analytical and simulated theoretical local spectra. The stellar disk is discretized by a 2D uniform square grid, each cell being associated with a specific local intensity spectrum. 

The simulated component of these intensity spectra is defined using the \textit{Turbospectrum code for spectral synthesis}\footnote{\url{https://github.com/bertrandplez/Turbospectrum2019}} \citep{plez2012}. This code uses MARCS photospheric models \citep{MARCS}\footnote{\url{https://marcs.astro.uu.se}} and spectral line-lists from VALD3 database\footnote{\url{http://vald.astro.uu.se}} \citep{Ryabchikova2015} to generate synthetic spectra under the assumption of local thermodynamic equilibrium\footnote{We used the \texttt{interpol$\_$marcs} module to derive a MARCS model for the exact values of temperature, metallicity and log \textit{g} of our target stars. Available for downloads at \url{https://marcs.astro.uu.se/software.php}}. For each star, we used \texttt{Turbospectrum } to generate high-resolution intensity spectra for a series of positions along the stellar radius to sample broadband limb-darkening and CLV. \\
These synthetic spectra, however, do not contain the He\,I triplet lines at 10830\AA\, as its formation in stellar atmospheres necessitates non-local thermodynamical equilibrium conditions that are usually met in chromospheric layers, whereas MARCS models focus on the photospheric layers. We thus calculate the He\,I triplet absorption lines analytically, assuming Gaussian cross-sections and a common temperature and density for the metastable helium gas. Figure \ref{fig:HAT_P_33_Imu} shows a series of intensity spectra across the stellar disk of HAT-P-33. \\

The series of synthetic+analytical intensity spectra is then interpolated over the whole stellar grid, and Doppler-shifted according to the local radial velocity set by the projected stellar rotational velocity \citep{Bourrier2022}. Subsequently, intensity spectra are scaled into local flux spectra using the surface of the stellar grid cells, and summed over the whole grid to derive a simulated disk-integrated spectrum of the target star. Finally, the disk-integrated spectra are convolved with GIANO-B instrumental response and resampled to match its spectral resolution. The observed and simulated disk-integrated spectra are compared using a MCMC fit with free parameters set to the temperature and density of the metastable helium atoms.\\

\begin{figure}
\includegraphics[ width =\linewidth]{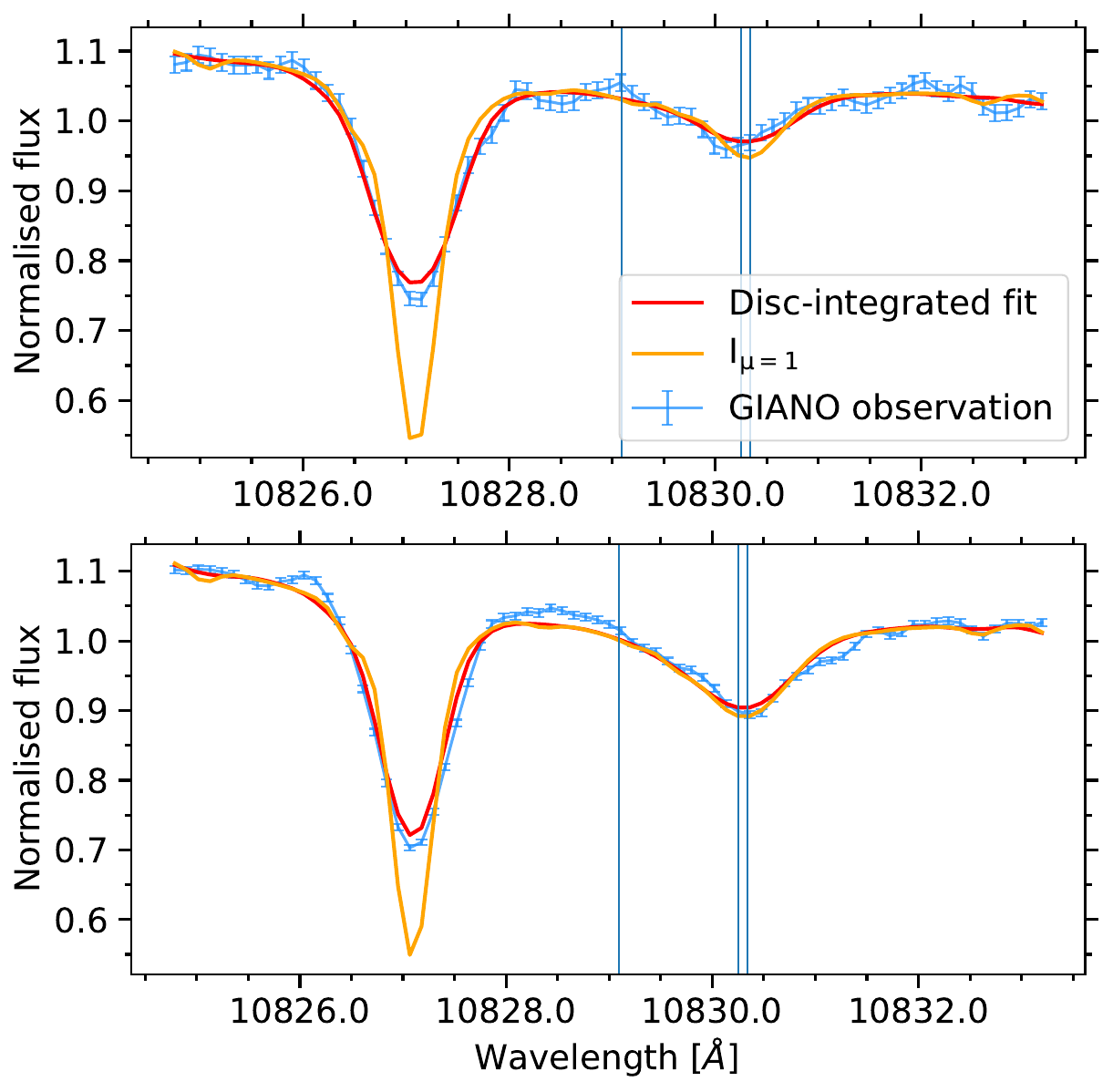}
\caption{Normalized spectra of HAT-P-33 (upper panel) and HAT-P-49 (lower panel). We show the best fit (red curve) to the observed (blue curve) disk-integrated spectrum, and the spectrum at the disk's center (orange curve). The blue vertical lines show the transition wavelength of the He\,I triplet.}
\label{fig:HATP_P_33_49_I0}
\end{figure}

Figure \ref{fig:HATP_P_33_49_I0} shows the results of our fits for HAT-P-33 and HAT-P-49, which have the highest $v$sin$i_\star$ of our sample, highlighting the local spectrum at disk's center that is later used in Eq. \ref{eq_T}. We only applied this approach to the four targets with the highest $v$sin$i_\star$, as POLDs are expected to be negligible for the other targets. For these slow rotators, the rotational broadening of the disk-integrated line profiles is small and they remain good proxies for the local planet-occulted lines, especially for the shallow He\,I triplet lines. We note that CLV is not accounted for in our analytical estimates of the He\,I lines.\\

We underline that even for the four fast-rotating targets for which we used a more accurate proxy for the planet-occulted stellar line the amplitude of the POLD was found to be comparable to the dispersion of the data. This is partly due to the shallowness of the He\,I triplet lines, and to the fact that POLDs partially smooth out when averaging transmission spectra in the planet rest frame over the transit window. Indeed, POLDs shift along the stellar surface RVs, while planetary signatures shift along the planet orbital RVs  (see Fig.~\ref{ros_dif}). Therefore, our final interpretation is made based on the transmission spectra calculated with the disk-integrated spectrum.

\section{Results}\label{results}
\begin{table*}[]
    \centering
    \caption{\ion{He}{I} measurements.}
    \small
    \begin{tabular}{c|c c c c c c c}
    \hline \hline
         \rule{0pt}{3ex} Target & Roche lobe & g$_{\mathrm{P}}$  & Excess of absorption & H$_{\mathrm{eq}}$ & $\delta_\mathrm{R_P}$/H$_{\mathrm{eq}}$ & F$_\mathrm{5-504~\textnormal{Å}}$ & $\dot{M}$~3$\sigma$ \\
         & [R$_\mathrm{P}$] &[m s$^{-2}$]& [$\%$] &[Km] & & [10$^{3}$ erg s$^{-1}$ cm$^{-2}$] &  [10$^{10}$ g~s$^{-1}$] \\
         \hline
HAT-P-3b& 3.647 & 17.5 & < 1.9 & 424 & 92 & 7.968 & 7.772\\
HAT-P-33b & 2.167 & 5.1 & < 1.4 & 2231 & 33 & 6.195 &  8.205\\
HAT-P-49b & 3.207  & 21.5 & < 0.6 & 629 & 61 & 14.51 &  5.702 \\
HD89345b & 8.213  & 7.8 & < 0.7 & 855 & 76 & 0.244 & 3.899 \\
K2-105b & 11.202  & 23.6 & < 2.33 & 219 & 378 & 14.69 &  6.028\\
Kepler-25c & 8.148  & 5.3 & < 1.86 & 1192 & 82 & 1.019 &  65.26\\
Kepler-68b & 8.722  & 15.2 & < 0.72 & 533 & 112 & 1.176 &  8.134\\
WASP-47d & 9.565  & 10.9 & < 3.29 & 533 & 230 & 0.577 &  7.854\\
    \end{tabular}
     \tablefoot{From left to right: Roche lobe (estimated following \citealt{Eggleton1983}), planet’s surface gravity, excess of absorption, atmospheric scale heights (computed by assuming $\mu$=1.3, Sect~\ref{results}), $\delta_\mathrm{R_P}$/H$_{\mathrm{eq}}$ (i.e. the 3$\sigma$ upper limits), the stellar XUV flux at the planet position, and the mass-loss rates computed with the \textit{p-winds} code at 3$\sigma$. }
    \label{table_result}
\end{table*}

\begin{figure*}
\includegraphics[ width =\textwidth]{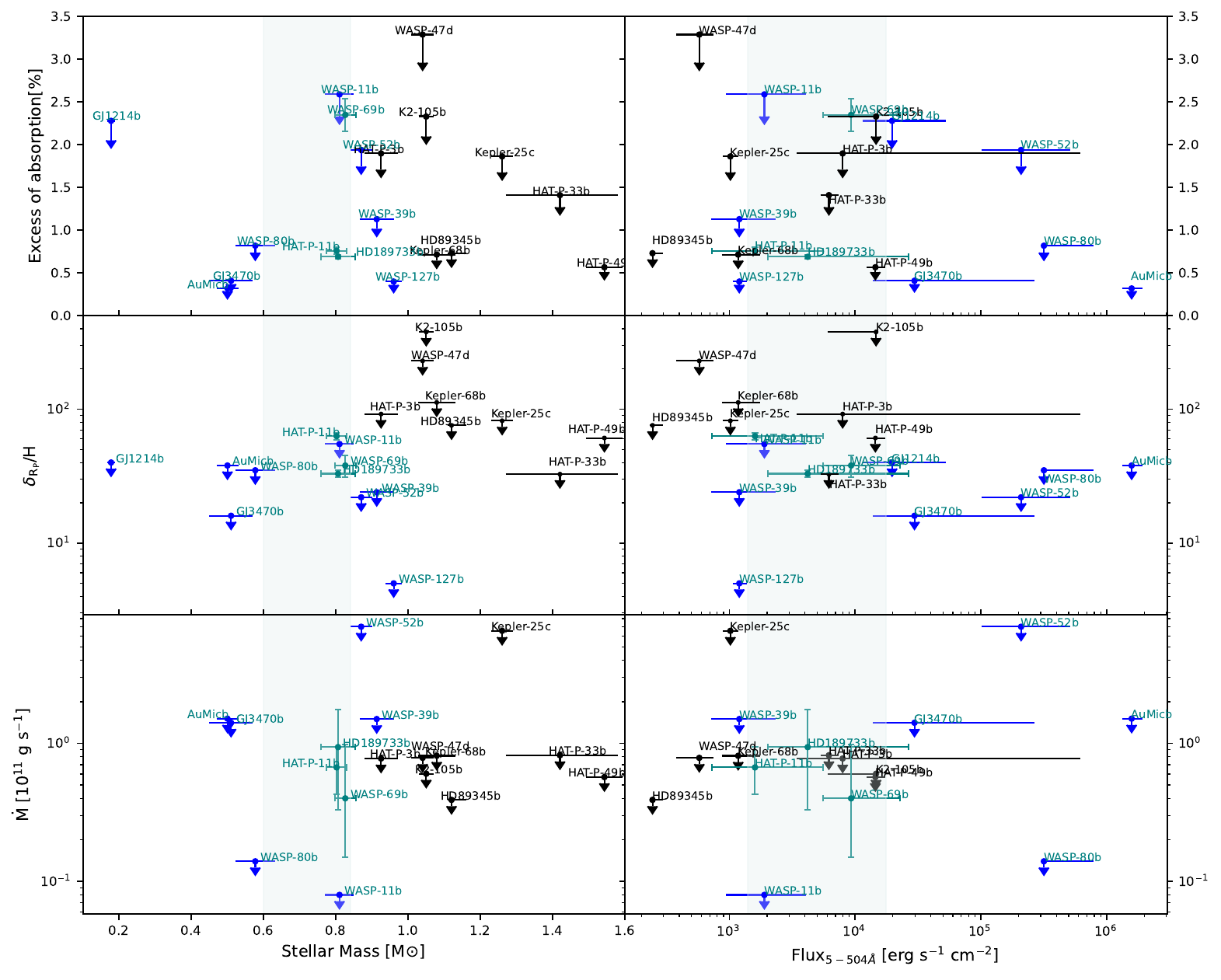}
\caption{Correlation plots. Excess of absorption (top panels), $\delta_\mathrm{R_P}$/H$_{\mathrm{eq}}$ (middle panels) and mass-loss rates (bottom panels) as a function of stellar mass and XUV irradiation at the planet position. Upper limits are given at 3$\sigma$. Blue stars are targets reported from \citet{Allart2023}, the detections and the non-detections are plotted in dark- and light-blue, respectively.  We note that K2-105~b observations suffered from GIANO-B's auto-guide problems and low S/N, so the limit we set on $\delta_\mathrm{R_P}/H_\mathrm{eq}$ is not constraining. We highlighted in gray the area of the parameter space that accordingly to \citet{Allart2023} seems to favor the detections. The errors on the XUV flux have been calculated with the error on the stellar ages.} 
\label{fig_result}  
\end{figure*}

The presence of an extended and possibly escaping helium atmosphere would appear as an absorption feature in the transmission spectrum in the planet’s rest frame at the position of the stellar helium triplet. Unfortunately, as shown in Fig.~\ref{tomo}, we did not detect significant helium absorption features for any of our targets. 
We thus evaluated 3$\sigma$ upper limits from the data itself as in \citet{Allart2023}.
Following an approach similar to that of \citet{Cubillos2017}, we computed Allan plots (see Fig.~\ref{Allan}) to estimate the  noise present in the data. We assumed the white noise $\sigma_1$ to be the standard deviation (hereafter, root mean square, rms) of the transmission spectrum, excluding the helium triplet region (1083.0-1083.6~nm).
 Then, we binned the transmission spectrum by bins of $N$ elements each and calculated the rms of the binned transmission spectrum. We repeat the process for a wide range of $N$ elements for bin (from 1 to 42). In the absence of correlated noise, $\sigma_1$ scales as $\sqrt{N}$.
 We then fitted the rms in a log-log space to derive the trend of the noise, the fitted rms at 0.075~nm is the 1~$\sigma$ uncertainty.  We set three times this value as the 3$\sigma$ upper limit on the signature contrast, $c$.\\
 An alternative approach to providing more rigorous estimations of the noise present in the data requires the use of Gaussian processes (See Appendix Sect.~\ref{GPL_sec}). However, for the rest of the analysis, to maintain consistency with the results published in \citet{Allart2023}, we used upper limits estimated from Allan plots.
\\
We then derived an upper limit on the equivalent opaque radius $\delta_\mathrm{R_P}$, namely, the height of an opaque atmospheric layer that would produce the observed absorption signal, as:
\begin{equation}
\delta_\mathrm{R_P}=\sqrt{(R_\mathrm{P}^2+R_\mathrm{\star}^2 \times c)}-R_\mathrm{P}\,
,\end{equation}

where $R_\mathrm{p}$ and ${R_\star}$ are the planetary and stellar radius, respectively.\\
We finally computed the quantity $\delta_\mathrm{R_P}/H_\mathrm{eq}$ \citep{Nortmann2018}, which expresses the number of scale heights ($H_\mathrm{eq}$) probed by the atmosphere in the considered spectral range, with $H_\mathrm{eq}$ = $\frac{k_\mathrm{B}T\mathrm{eq}}{\mu g}$ and $k_\mathrm{B}$ the Boltzmann constant, $T_\mathrm{eq}$ the planetary equilibrium temperature (listed in Table~\ref{Tab_parameters}), $g$ the planetary gravity computed from the planetary mass and radius (reported in Table~\ref{table_result}), $\mu$ the mean molecular weight (for which we assumed a hydrogen-dominated atmosphere and hence a value of 1.3 times the mass of a hydrogen atom). Table~\ref{table_result} reports the derived $\delta_\mathrm{R_P}/H_\mathrm{eq}$ values for each investigated planet. \\

We explored how the derived constraints vary as a function of the stellar mass and XUV flux between 5 and 504 \AA, which are the energies mainly responsible for the population of the metastable \ion{He}{I} level \citep{sanz-forcada2011}\footnote{For consistency, in Appendix in Fig.~\ref{mid_flux} we reported also the same plots as a function of the insolation level of mid-UV flux, which accordingly to \citealt{Oklop2019} ionizes the Helium's metastable state}. We focused on these two parameters because \citet{Allart2023} showed that they do yield visible trends with $\delta_\mathrm{R_P}/H_\mathrm{eq}$. Trends related to the excess of absorption and atmospheric extension are shown in the top and middle panels of Fig.~\ref{fig_result}. All our targets are outside the area of the parameter space, with masses between $\sim$0.6 and $\sim$0.85~M$_\sun$, as pointed out by \citet{Allart2023} to favor the \ion{He}{I} detection. This range of stellar masses corresponds to K-stars, in agreement with the predictions of \citet{Oklop2019}, thus our non-detections are not entirely unexpected. 
However, half of our non-detections do not agree with the XUV flux range found to favor the presence of \ion{He}{I} in \citet{Allart2023}, 1400-17800~erg$\,$s$^{-1}\,$cm$^{-2}$.  However, the XUV flux values depend on the model and are associated with star ages, which are typically not well-constrained. So we have to be cautious with these values. Finally, as already highlighted in \citet{Allart2023}, even if surprising, there are no clear correlations between \.M and the stellar mass or XUV flux (bottom panels of Fig.~\ref{fig_result}).\\

We chose to focus on the targets analyzed in this paper and in \citet{Allart2023}, rather than  other results from the literature because we derived the absorption, atmospheric extension, and mass-loss values homogeneously with the same methodology. \\

\section{Summary and conclusions} \label{summary}
In this paper, we describe our  \ion{He}{I} survey of 9 planets at the edge of the Neptunian desert, with the goal of understanding the role of photoevaporation in sculpturing this feature.
We analyzed observations gathered with the high-resolution GIANO-B spectrograph mounted on the TNG, and we used the transmission spectroscopy technique to detect a possible extended or evaporating helium atmosphere in the investigated planets. We found no sign of planetary absorption at the position of the stellar \ion{He}{I} triplet in any of the investigated targets, and we thus provided 3$\sigma$ upper limits on the \ion{He}{I} absorption. 
We underline that the GIANO-B transmission spectra are affected by various systematics that are not fully understood and difficult to properly remove. These systematics may be caused by low data quality (e.g., low S/N) or instrumental effects (e.g., auto-guide problems).\\

We interpreted our derived transmission spectra with the \textit{p-wind} code \citep{dossantos2022} and we attempted to interpret our findings by putting them in the wider context of the measurements presented in Allart et al. 2023 submitted. We searched for correlations with the stellar mass and XUV flux \citep[e.g.,][]{Fossati2022}, which are thought to be key drivers in the formation of the \ion{He}{I} triplet. 
Constraints from our sample support the trend of $\delta_\mathrm{R_P}/H_\mathrm{eq}$, with the stellar mass proposed by \citet{Allart2023}, which remains a good indicator for the presence of metastable helium in exoplanet atmospheres. In addition, they are not incompatible with the trend highlighted in \citet{Allart2023} with the XUV flux as they are not constraining enough to reach a better precision. We stress the importance of carrying out helium surveys with the same instrument and analyzing them with the same data reduction technique, as heterogeneity can obscure any trends in the data \citep{Vissapragada2022}. Several instruments are now available to perform this kind of homogeneous survey such as NIRSPEC, SPIROU, CARMENES, and GIANO-B.

\bibliographystyle{aa}
\bibliography{ref}
\begin{acknowledgements}
        We thank the referee for the comments and suggestions. We thank V.l Andretta for his help. G.G. acknowledge financial contributions from PRIN INAF 2019, and from the agreement ASI-INAF number 2018-16-HH.0 (THE StellaR PAth project). 
        R. A. is a Trottier Postdoctoral Fellow and acknowledges support from the Trottier Family Foundation. This work was supported in part by a grant from the Fonds de Recherche du Québec - Nature et Technologies (FRQNT). This work was funded by the Institut Trottier de Recherche sur les Exoplaneètes (iREx).  
        This work has been carried out within the framework of the NCCR PlanetS supported by the Swiss National Sci- ence Foundation under grants 51NF40182901 and 51NF40205606. This project has received funding from the European Research Council (ERC) under the Eu- ropean Union’s Horizon 2020 research and innovation programme (project Spice Dune, grant agreement No 947634). This material reflects only the authors views and the Commission is not liable for any use that may be made of the information contained therein.
        This work has made use of the VALD database, operated at Uppsala University, the Institute of Astronomy RAS in Moscow, and the University of Vienna. This work has made use of the \textit{Turbospectrum code for spectral synthesis}
\end{acknowledgements}
\begin{appendix}
        
        \section{Additional figures and tables} \label{add}
        \begin{table}[h]
                \begin{minipage}{\textwidth}
                        
                        \caption{Planets with \ion{He}{I} study reported in the literature. }
                        \label{det_he}
                        \centering  
                        \begin{tabular}{l | c |  l }   
                                \hline\hline                       
                                Planet & Status & Reference \\  
                                \hline  
                                WASP-11b & \xmark &\citealt{Allart2023}\\
                                WASP-12b & \xmark &\citealt{Kreidberg2018}\\
                                WASP-39b & \xmark &\citealt{Allart2023}\\
                                WASP-48b & \xmark &\citealt{Bennett2023}\\
                                WASP-52b & $\sim$ &\citealt{Vissapragada2020b,Kirk2022,Allart2023}\\
                                WASP-69b & \cmark &\citealt[][]{Nortmann2018, Vissapragada2020b,Allart2023}\\ 
                                WASP-80b & \xmark &\citealt{Fossati2022,Vissapragada2022,Allart2023}\\
                                WASP-76b & \xmark &\citealt{Casasayas-Barris2021}\\
                                WASP-107b & \cmark & \citealt[][]{Spake2018, Allart2019, kirk2020}\\
                                WASP-127b & \xmark &\citealt{dosSantos2020,Allart2023}\\
                                WASP-177b & \xmark &\citealt{Kirk2022,Vissapragada2022}\\
                                HD189733b & \cmark &\citealt{Salz2018,Guilluy2020,Zhang2022,Allart2023}\\ 
                                HD209458b & $\sim$ &\citealt{Nortmann2018,Alonso-Floriano2019}\\
                                HD97658b & \xmark &\citealt{Kasper2020}\\
                                HD63433b & \xmark &\citealt{Zhang2022a}\\
                                HD63433c & \xmark &\citealt{Zhang2022a}\\
                                HAT-P-11b & \cmark &\citealt{Allart2018, Mansfield2018,Allart2023}\\ 
                                HAT-P-18b & \cmark &\citealt{Paragas2021,Vissapragada2022}\\
                                HAT-P-26b & \cmark &\citealt{Vissapragada2022}\\
                                HAT-P-32b & \cmark &\citealt{Czesla2022}\\
                                NGTS-5b & $\sim$ &\citealt{Vissapragada2022}\\
                                55Cnce & \xmark &\citealt{Zhang2021}\\
                                Kelt-9b & \xmark &\citealt{Nortmann2018}\\
                                GJ3470b & $\sim$ &\citealt{ Ninan2020,Allart2023,Palle2020}\\ 
                                GJ436b & \xmark  & \citealt{Nortmann2018}\\
                                GJ1214b & $\sim$ &e.g.,\citealt[][]{Orell-Miquel2022,Kasper2020,Allart2023}\\
                                GJ9827b & \xmark &\citealt{Carleo2021,Krishnamurthy2023}\\
                                GJ9827d & \xmark &\citealt{Kasper2020,Carleo2021,Krishnamurthy2023}\\
                                TOI-560b  & \cmark &\citealt{Zhang2022,Zhang2022b}\\
                                TOI-1728b &\xmark & \citealt{Kanodia2020}\\
                                TOI-1430b & \cmark &\citealt{Zhang2022b}\\
                                TOI-1683b & \cmark &\citealt{Zhang2022b}\\
                                TOI-1807b & \xmark & \citealt{Gaidos2022}\\
                                TOI-2076b & \xmark &\citealt{Zhang2022b,Gaidos2022}\\
                                TOI-3757b & \xmark & \citealt{Kanodia2022}\\
                                TOI-1235b & \xmark & \citealt{Krishnamurthy2023}\\
                                TOI-2136b & \xmark & \citealt{Kawauchi2022}\\
                                Trappist-1b & \xmark & \citealt{Krishnamurthy2021}\\
                                Trappist-1e & \xmark &  \citealt{Krishnamurthy2021}\\
                                Trappist-1f & \xmark & \citealt{Krishnamurthy2021}\\
                                V1298Taub &  $\sim$ & e.g., \citealt{Vissapragadav1298,Gaidos2022_v1298}\\
                                V1298Tauc & \xmark & e.g., \citealt{Vissapragadav1298}\\
                                V1298Taud & \cmark & e.g., \citealt{Vissapragadav1298} \\
                                AU Mic b & \xmark & \citealt{Hirano2020}\\
                                K2-25b & \xmark &  \citealt{Gaidos2020MNRAS}\\
                                K2-100b & \xmark & \citealt{Gaidos2020} \\
                        \end{tabular}
                        \tablefoot{From left to right: the investigated planet, the status of the detection (with \cmark, \xmark, and $\sim$ indicating detections, non-detections, and not-clear results respectively), and references.}
                \end{minipage}
        \end{table}
        
        \onecolumn
\begin{longtable}{lcc}
        \caption{Adopted parameters}    \\
        \hline\hline
        \small
        Parameter &Value & Reference \\ 
        \hline
        \endfirsthead
        \caption{continued}\\
        \hline
        Parameter &Value & Reference \\ 
        \hline
        \endhead
        \hline
        \endfoot
        \hline
        \endlastfoot
        
        \textbf{ HAT-P-3}  & & \\
        
        $\bullet$       \underline{\textit{Stellar parameters}} &  & \\
        
        Spectral type & K1 & \citet{Grieves2018} \\
        Stellar mass, M$_\star$ (M$_\sun$) & 0.925 $\pm$ 0.046  & \citet{Mancini2018A}\\
        Stellar radius, R$_\star$ (M$_\sun$) & 0.850 $\pm$ 0.021 & \citet{Mancini2018A}\\
        Stellar age, $\tau$ (Gyr) & $2.9^{+2.7}_{-4.9}$ & \citet{Mancini2018A}\\
        Effective temperature, Teff (K) & 5190  $\pm$ 80 & \citet{Mancini2018A} \\
        Metallicity (dex)  &  0.24 $\pm$ 0.08 (Fe/H) & \citet{Mancini2018A} \\
        log~g (log$_{10}$(\cms)) & 4.545 $\pm$ 0.023  &  \citet{Mancini2018A} \\
        Systemic velocity, v$_\mathrm{sys}$ (\kms ) & -23.379680  & DREAM I\\
        Limb-darkening coefficients $\mu_1$& 0.216     & EXOFAST\footnotemark[1] \\
        $\,\,\,\,\,\,\,\,\,\,\,\,\,\,\,\,\,\,\,\,\,\,\,\,\,\,\,\,\,\,\,\,\,\,\,\,\,\,\,\,\,\,\,\,\,\,\,\,\,\,\,\,\,\,\,\,\,\,\,\,\,\,\,\,\,\,\,\,\,\,\,$ $\mu_2$ & 0.286 & EXOFAST\footnotemark[1] \\
        Stellar projected velocity, $v$sin$i_\star$ (\kms) & $0.46^{+0.22}_{-0.25}$  & DREAM 1 \\
        Magnitude (J-band) & 9.936$\pm$0.022 & \citet{Vizier}\\
        $\bullet$       \underline{\textit{Planetary parameters}} & & \\
        Orbital period, P (days) & 2.89973797 $\pm$ 0.00000038& \citet{Baluev2019}\\
        Transit epoch, T$_0$ (BJD$_\mathrm{TDB}$) & 2454218.75960 $\pm$  0.00016 & DREAM I  \\
        Eccentricity, e & 0.0 (fixed) & \citet{Mancini2018A} \\
        Argument of periastron, $\omega_\star$ & 90 (fixed) & \citet{Mancini2018A} \\
        Stellar reflex velocity, K$_\star$ (\ms) & 90.63 $\pm$ 0.58  & \citet{Mancini2018A} \\
        Scaled separation, a/R$_\star$ & 9.8105 $\pm$ 0.2667 & \citet{Mancini2018A}\\
        Orbital inclination, $i$ & 86.31 $\pm$ 0.19 deg & \citet{Mancini2018A} \\
        Planet-to-star radius ratio, R$_\mathrm{P}$/R$_\star$ &0.11091 $\pm$ 0.00048   & \citet{Baluev2019}\\
        Planetary mass, M$_\mathrm{pl}$ (\mjup) &  0.595 $\pm$0.019  & \citet{Mancini2018A}\\
        Planetary density,  $\rho_\mathrm{pl}$ (\gcm) & 0.9750$\pm$0.1000  &      \citet{Mancini2018A}\\
        Projected spin-orbit angle, $\lambda$ (deg) & -25.3$^{+29.4}_{ - 22.8}$  & DREAM I \\
        Planet radial-velocity semi-amplitude, K$_\mathrm{pl}$(\kms) & 145.2$\pm$2.4 & This paper\footnotemark[2] \\
        Equilibrium temperature, T$_\mathrm{eq}$(K) & 1170$\pm$17 & 
        \citet{Mancini2018A}\\
        & &  \\
        \hline
        \textbf{ HAT-P-33}  & & \\
        
        $\bullet$       \underline{\textit{Stellar parameters}} &  & \\
        Spectral type & F4 & \citet{Luo2018} \\
        Stellar mass, M$_\star$ (M$_\sun$) &1.42 $^{+0.16}_{-0.15}$  & \citet{Wang2017}   \\
        Stellar radius, R$_\star$ (M$_\sun$) & 1.91$^{+0.26}_{-0.20}$ & \citet{Wang2017}\\
        Stellar age, $\tau$ (Gyr) & 2.30 $\pm$ 0.30 & \citet{bonomo2017}\\
        Effective temperature, teff (K) &6460 $^{+300}_{ -290} $ & \citet{Wang2017}\\
        Metallicity  (dex)&  0.01 $\pm$  0.31 [Fe/H] & \citet{Wang2017}\\
        Surface gravity, log g$_\star$ (cgs) &   $4.030 _{-0.090}^{+0.079}$  & \citet{Wang2017} \\
        Systemic velocity, v$_\mathrm{sys}$ (\kms) & 23.080601  & DREAM I\\
        Limb-darkening coefficients $\mu_1$& 0.097      & EXOFAST\footnotemark[1] \\
        $\,\,\,\,\,\,\,\,\,\,\,\,\,\,\,\,\,\,\,\,\,\,\,\,\,\,\,\,\,\,\,\,\,\,\,\,\,\,\,\,\,\,\,\,\,\,\,\,\,\,\,\,\,\,\,\,\,\,\,\,\,\,\,\,\,\,\,\,\,\,\,$ $\mu_2$  & 0.301  & EXOFAST\footnotemark[1] \\
        Stellar projected velocity, $v$sin$i_\star$ (\kms)& 15.57 $\pm$ 0.31  & DREAM I\\
        Magnitude (J-band) & 10.263$\pm$0.021 & \citet{Vizier}\\
        $\bullet$       \underline{\textit{Planetary parameters}} & & \\
        Orbital period, P (days) & 3.47447773 $\pm$ 0.00000066 &  DREAM I \\
        Transit epoch, T$_\mathrm{0}$ (BJD$_\mathrm{TDB}$) & 2456684.86508 $\pm$ 0.00027   & DREAM I \\
        Eccentricity, e &  0.180 $_{-0.096} ^{+0.110}$&  DREAM I\\
        Argument of periastron, $\omega_\star$ (deg)& 88$^{+33}_{ -34}$ & \citet{Wang2017} \\
        Stellar reflex velocity, K$_\star$ \ms & 74.4$\pm$8.5 &  DREAM I \\
        Scaled separation, a/R$_\star$ & 5.69$^{+0.58}_{-0.59}$&  \citet{Wang2017} \\
        Orbital inclination, $i$ (deg) &  88.2 $^{+1.2}_{ -1.3}$  & \citet{Wang2017}   \\
        Planet-to-star radius ratio, R$_\mathrm{P}$/R$_\star$ & 0.10097 $^{+0.00056} _{-0.00052}$ & \citet{Wang2017} \\
        Planetary mass, M$_\mathrm{pl}$ (\mjup)  & 0.72 $^{+0.13}_{ -0.12}$  & \citet{Wang2017} \\
        Planetary density, $\rho_\mathrm{pl}$  (\gcm) & 0.134$_{-0.042}^{+0.053}$  &      \citet{Wang2017}\\
        Projected spin-orbit angle, $\lambda$ (deg) & 5.9$\pm$4.1 deg & DREAM I \\
        Planet radial velocity semi-amplitude, K$_\mathrm{pl}$(\kms) & 160.6$^{+6.9}_{-6.3}$ & This paper\footnotemark[2] \\
        Equilibrium temperature, T$_\mathrm{eq}$(K) & 1782$\pm$28 & 
        \citet{Hartman2011}\\
        & & \\
        \hline
        \textbf{ HAT-P-49}  & & \\
        
        $\bullet$       \underline{\textit{Stellar parameters}} &  & \\
        Spectral type & F3 & DREAM I \\
        Stellar mass, M$_\star$ (M$_\sun$) & 1.543$\pm$0.051 & \citet{Bieryla2014}\\
        Stellar radius, R$_\star$ (R$_\sun$) & $1.833{+0.138}_{-0.076}$ & \citet{Bieryla2014}\\
        Stellar age $\tau$ (Gyr) & 1.50 $\pm$ 0.20 & \citet{bonomo2017}\\
        Effective temperature, Teff  (K) & 6820$\pm$52& \citet{Bieryla2014}\\
        Metallicity (dex) & 0.074$\pm$0.080  [Fe/H]& \citet{Bieryla2014}\\
        Surface gravity, log g$_\star$ (cgs) & 4.10$\pm$0.04   & \citet{Bieryla2014} \\
        Systemic velocity, v$_\mathrm{sys}$ (\kms)& 14.208478  & DREAM I\\
        Limb-darkening coeffcients $\mu_1$&   0.078         & EXOFAST\footnotemark[1] \\
        $\,\,\,\,\,\,\,\,\,\,\,\,\,\,\,\,\,\,\,\,\,\,\,\,\,\,\,\,\,\,\,\,\,\,\,\,\,\,\,\,\,\,\,\,\,\,\,\,\,\,\,\,\,\,\,\,\,\,\,\,\,\,\,\,\,\,\,\,\,\,\,$ $\mu_2$  & 0.303  & EXOFAST\footnotemark[1] \\
        Stellar projected velocity, $v$sin$i_\star$ (\kms) & 10.68$^{+0.46}_{ -0.47}$  & DREAM I\\
        Magnitude (J-band) & 9.550$\pm$0.020 & \citet{Vizier}\\
        $\bullet$       \underline{\textit{Planetary parameters}} & & \\
        Orbital period, P (days) & 2.6915539$\pm$0.0000012 &  DREAM I \\
        Transit epoch, T$_\mathrm{0}$ (BJD$_\mathrm{TDB}$) (BJD$_\mathrm{TDB}$) & 2456975.61736 $\pm$0.00050   & DREAM I \\
        Eccentricity, e & 0.0 (fixed) & \citet{Bieryla2014} \\
        Argument of periastron, $\omega_\star$ (deg) & 90 (fixed)  & \citet{Bieryla2014}\\
        Stellar reflex velocity, K$_\star$ (\ms)& 177.6 $\pm$ 16.0 & DREAM I \\
        Scaled separation, a/R$_\star$ & 5.13 $^{+0.19}_{ -0.30}$ & \citet{Bieryla2014}\\
        Orbital inclination, $i$ (deg) & 86.2$\pm$1.7 & \citet{Bieryla2014}\\
        Planet-to-star radius ratio, R$_\mathrm{P}$/R$_\star$ & 0.0792 $\pm$  0.0019 & \citet{Bieryla2014} \\
        Planetary M$_\mathrm{pl}$ (\mjup)& 1.730 $\pm$  0.205 & \citet{Bieryla2014}\\
        Planetary density, $\rho_\mathrm{pl}$ (\gcm) & 0.75$\pm$0.17  &  \citet{Bieryla2014}\\
        Projected spin-orbit angle, $\lambda$ (deg)& -97.7$\pm$1.8 & DREAM I \\
        Planet radial velocity semi-amplitude, K$_\mathrm{pl}$(\kms) & 176.5$\pm$2.0 & This paper\footnotemark[2] \\
        Equilibrium temperature, T$_\mathrm{eq}$(K) & 2131$^{+69}_{ -42}$ & \citet{Bieryla2014} \\
        & & \\
        \hline
        \textbf{HD89345}  & & \\
        
        $\bullet$       \underline{\textit{Stellar parameters}} &  & \\
        Spectral type & G5 &\citet{Cannon1993} \\
        Stellar mass, M$_\star$ (M$_\sun$) & 1.120$^{+0.040}_{ -0.010}$   & \citet{Van2018}\\
        Stellar radius, R$_\star$ (R$_\sun$) & 1.657$^{+0.020}_{ -0.004}$   & \citet{Van2018}\\
        Stellar age, $\tau$ (Gyr) & 9.40 $^{+0.40}_{-1.30}$ & \citet{Van2018}\\
        Effective temperature, Teff (K) & 5499  $\pm$ 73 &  \citet{Van2018}\\
        Metallicity  (dex) & 0.45$\pm$ 0.04 [Fe/H]& \citet{Van2018}\\
        Surface gravity log~g$_\star$ (log$_{10}$(\cms))& 4.044$^{+0.006}_{ -0.004}$ & \citet{Van2018}\\
        Systemic velocity, v$_\mathrm{sys}$ (\kms) &  2.223394  & DREAM I\\
        Limb-darkening coeffcients $\mu_1$&    0.182           & EXOFAST\footnotemark[1]\footnotemark[2] \\
        $\,\,\,\,\,\,\,\,\,\,\,\,\,\,\,\,\,\,\,\,\,\,\,\,\,\,\,\,\,\,\,\,\,\,\,\,\,\,\,\,\,\,\,\,\,\,\,\,\,\,\,\,\,\,\,\,\,\,\,\,\,\,\,\,\,\,\,\,\,\,\,$ $\mu_2$  &  0.300  & EXOFAST\footnotemark[1] \\
        Stellar projected velocity, $v$sin$i_\star$ (\kms) & 0.58$\pm$0.28  & DREAM I \\
        Magnitude (J-band) &  8.091 $\pm$0.020  & \citet{Vizier}\\
        $\bullet$       \underline{\textit{Planetary parameters}} & & \\
        Orbital period, P (Days) &  11.8144024 $\pm$0.0000066&  DREAM I \\
        Transit epoch T$_\mathrm{0}$ (BJD$_\mathrm{TDB}$)& 2458740.81147$\pm$0.00044   & DREAM I \\
        Eccentricity, e & 0.208 $\pm$ 0.039 &  DREAM I \\
        Argument of periastron $\omega$ (deg)& 21.7 $\pm$  19.1 & DREAM I  \\
        Stellar reflex velocity, K$_\star$ (\ms)& 9.1 $\pm$ 0.5 &  DREAM I \\
        Scaled separation, a/R$_\star$ & 13.625 $\pm$ 0.027& \citet{Van2018}\\
        Orbital inclination, $i$ (deg) & 87.68 $\pm$  0.10 & DREAM I  \\
        Planet-to-star radius ratio, R$_\mathrm{P}$/R$_\star$ & 0.03696 $\pm$ 0.00041 & DREAM I \\
        Planetary mass, M$_\mathrm{pl}$ (\mjup)& 0.112$\pm$0.010  &      \citet{Van2018}\\
        Planetary density, $\rho_\mathrm{pl}$ (\gcm) &  0.609$\pm$0.067 &  \citet{Van2018}\\
        Projected spin-orbit angle, $\lambda$ (deg) & 74.2$^{+33.6}_{ -32.5}$  & DREAM 1 \\
        Planet radial velocity semi-amplitude, K$_\mathrm{pl}$(\kms) & 99.2$^{+1.4}_{ -0.9}$ & This paper\footnotemark[2] \\
        Equilibrium temperature, T$_\mathrm{eq}$(K) & 1053$\pm$14 & \citet{Van2018}\\
        & &  \\
        \hline
        \textbf{ K2-105}  & & \\
        
        $\bullet$       \underline{\textit{Stellar parameters}} &  & \\

        Spectral type & G5 & This paper\footnotemark[5]\\
        Stellar mass,   M$_\star$ (M$_\sun$)&    1.05 $\pm$ 0.02   & \citet{Castro2022}  \\
        Stellar radius,   R$_\star$ (R$_\sun$)&  0.97 $\pm$ 0.01   & \citet{Castro2022}  \\
        Stellar age, $\tau$ (Gyr) & >0.6 & \citet{Narita2017}\\
        Effective temperature, Teff (K) & 5636$^{+49}_{ -52}$ & \citet{Castro2022} \\
        Metallicity (dex)& 0.23$^{+0.04}_{ -0.03}$ [Fe/H]  & \citet{Castro2022} \\
        Surface gravity log~g$_\star$ (log$_{10}$(\cms))&       4.49 $\pm$ 0.01  & \citet{Castro2022} \\
        Systemic velocity,  v$_\mathrm{sys}$ (\kms)&  -32.390637 & DREAM I \\
        Limb-darkening coefficients $\mu_1$ & 0.169     & EXOFAST\footnotemark[1] \\
        $\,\,\,\,\,\,\,\,\,\,\,\,\,\,\,\,\,\,\,\,\,\,\,\,\,\,\,\,\,\,\,\,\,\,\,\,\,\,\,\,\,\,\,\,\,\,\,\,\,\,\,\,\,\,\,\,\,\,\,\,\,\,\,\,\,\,\,\,\,\,\,$ $\mu_2$  &      0.299 & EXOFAST\footnotemark[1] \\
        Stellar projected velocity, $v$sin$i_\star$ (\kms)& 2.13$^{+0.96}_{-0.92}$  & DREAM I \\
        Magnitude (J-band) & 10.541$\pm$0.02  & \citet{Vizier}\\
        $\bullet$       \underline{\textit{Planetary parameters}} & & \\
        Orbital period, P (days) & 8.2669897$\pm$0.0000057 &  DREAM I \\
        Transit epoch, T$_\mathrm{0}$ (BJD$_\mathrm{TDB}$) & 2458363.2387$^{+0.00069}_{ -0.000633}$ & DREAM I \\
        Eccentricity, e &  0 (fixed)&  DREAM I \\
        Argument of periastron, $\omega$ (deg)& 90 (fixed) &  DREAM I \\
        Stellar reflex velocity, K$_\star$ (\ms)& 9.4 $\pm$ 5.8 &  \citet{Narita2017}\\
        Scaled separation, a/R$_\star$ & 17.39 $\pm$  0.19 &  DREAM I \\
        Orbital inclination, $i$ (deg)&  88.62 $\pm$  0.10 & DREAM I    \\
        Planet-to-star radius ratio, R$_\mathrm{P}$/R$_\star$ & 0.03332 $\pm$  0.00067 & DREAM I    \\
        Planetary mass, M$_\mathrm{pl}$ (\mjup)& 0.094 $\pm$   0.060 & \citet{Narita2017}  \\
        Planetary density, $\rho_\mathrm{pl}$  (\gcm) & 2.3$^{+1.7}_{-1.6}$  &      This paper \\
        Projected spin-orbit angle, (deg) $\lambda$& -81$^{+50}_{-47}$ & DREAM I \\
        Planet radial-velocity semi-amplitude, K$_\mathrm{pl}$(\kms) & 107.0$\pm$0.7 & This paper\footnotemark[2] \\
        Equilibrium temperature, T$_\mathrm{eq}$(K) & 814$\pm$12 & \citet{Livingston2018} \\
        & & \\
        \hline
        \textbf{ Kepler-25}  & & \\
        $\bullet$       \underline{\textit{Stellar Parameters}} &  & \\
        
        Spectral type & F8 & DREAM I \\
        Stellar mass,   M$_\star$ (M$_\sun$) &   1.26 $\pm$ 0.03   & \citet{Benomar2014}  \\
        Stellar radius,   R$_\star$ (R$_\sun$) & 1.34$\pm$0.01& \citet{Benomar2014}  \\
        Stellar age, $\tau$ (Gyr) & 2.75 $\pm$ 0.30 & \citet{Benomar2014}\\
        Effective temperature, Teff (K) & 6354$\pm$27 &  \citet{Benomar2014} \\
        Metallicity [Fe/H] (dex) & 0.11$\pm$0.03   & \citet{Benomar2014}  \\
        Surface gravity log~g (log$_{10}$(\cms))& 4.285$\pm$0.003& \citet{Benomar2014}  \\
        Systemic velocity,  v$_\mathrm{sys}$ (\kms) & -8.633258 & DREAM I \\
        Limb-darkening coeffcients $\mu_1$ &  0.106      & EXOFAST\footnotemark[1] \\
        $\,\,\,\,\,\,\,\,\,\,\,\,\,\,\,\,\,\,\,\,\,\,\,\,\,\,\,\,\,\,\,\,\,\,\,\,\,\,\,\,\,\,\,\,\,\,\,\,\,\,\,\,\,\,\,\,\,\,\,\,\,\,\,\,\,\,\,\,\,\,\,$ $\mu_2$  &         0.304 & EXOFAST\footnotemark[1] \\
        Stellar projected velocity, $v$sin$i_\star$ (\kms)& 8.89$^{+0.59}_{ -0.63 }$ & DREAM I\\
        $\bullet$       \underline{\textit{Planetary parameters}} & & \\
        Magnitude (J-band) & 9.764 $\pm$0.020  & \citet{Vizier}\\
        $\star$ Planet, $b$  & & \\
        Orbital period, P (days) & 6.2385347882 & \citet{Battley2021}\\
        Transit epoch, T$_\mathrm{0}$ (BJD$_\mathrm{TDB}$) &  2458648.00807$^{+0.00057}_{ - 0.00051 }$ &DREAM I \\
        Eccentricity, e & 0.0029 $^{0.0023}_{ - 0.0017 }$&  \citet{Mills2019}\\\
        $\sqrt{e}\,\cos{\omega}$ & 0.042 $^{0.017}_{ - 0.036 }$ & \citet{Mills2019} \\
        $\sqrt{e}\,\sin{\omega}$ & 0.007 $^{0.038}_{ - 0.035 }$ & \citet{Mills2019} \\
        Stellar reflex velocity K$_\star$ (\ms) & 2.6$\pm$0.7   & This paper\footnotemark[3] \\
        Orbital inclination $i$ (deg)& 87.173$^{0.084}_{ - 0.083 }$ & \citet{Mills2019}\\\
        Planet-to-star radius ratio R$_\mathrm{P}$/R$_\star$ & 0.019160 $^{+5.1e-5}_{ -4.8e-5 }$ &  \citet{Mills2019}\\\
        Planetary M$_\mathrm{pl}$ & 0.0275 $^{0.0079}_{ - 0.0073 }$&  \citet{Mills2019}\\
        $\star$ Planet $c$  & & \\
        Orbital period P (days)& 12.720370495 $\pm$ 0.000001703 & \citet{Battley2021}\\
        Transit epoch T$_\mathrm{0}$ (BJD$_\mathrm{TDB}$) &  2458649.55482 $ _{-0.00051}^{ +0.00057}$ BJD$_\mathrm{TDB}$ &DREAM I \\
        Eccentricity e & 0.0061  $^{+0.0049}_{ - 0.0041}$ & \citet{Mills2019}\ \\
        $\sqrt{e}\,\cos{\omega}$ & -0.024 $^{0.067}_{ - 0.053 }$ & \citet{Mills2019} \\
        $\sqrt{e}\,\sin{\omega}$ & 0.004 $^{0.065}_{ - 0.062 }$ & \citet{Mills2019} \\
        Stellar reflex velocity, K$_\star$ (\ms) & 3.6$^{+0.3}_{-0.4}$ & This paper\footnotemark[3] \\
        Scaled separation, a/R$_\star$ & 18.336 $\pm$  0.27  & \citet{Mills2019}\\
        Orbital inclination, $i$ (deg)& 87.236 $ _{-0.039}^{ +0.042}$ &  \citet{Mills2019}\\\
        Planet-to-star radius ratio, R$_\mathrm{P}$/R$_\star$ & 0.03637 $\pm$  0.00012 &  \citet{Mills2019}\\\
        Planetary mass, M$_\mathrm{pl}$ (\mjup) & 0.0479  $^{+0.0041}_{ -   0.0051}$ &  \citet{Mills2019}\\
        Planetary density & 0.588$^{+0.053}_{-0.061}$ \gcm & \citet{Mills2019} \\
        Projected spin-orbit angle, $\lambda$& -0.9$^{+7.7}_{ - 6.4}$ & DREAM I \\
        Planet radial-velocity semi-amplitude, K$_\mathrm{pl}$(\kms) & 98.4$\pm$0.8 & This EXOFAST\footnotemark[2] \\
        Equilibrium temperature, T$_\mathrm{eq}$(K) & 992$\pm$8 & This work\footnotemark[4] \\
        $\star$ Planet, $d$  & & \\
        Orbital period, P (days) &  122.40$^{+0.80}_{ - 0.71 }$  d&  \citet{Mills2019}\ \\
        Transit epoch, T$_\mathrm{0}$ (BJD$_\mathrm{TDB}$) & 2455715.0$^{+6.8}_{ - 7.2 }$ &DREAM I \\
        Eccentricity, e & 0.13$^{+0.13}_{ - 0.09 }$  & \citet{Mills2019}\ \\
        $\sqrt{e}\,\cos{\omega}$ & 0.07 $^{0.027}_{ - 0.029 }$ & \citet{Mills2019} \\
        $\sqrt{e}\,\sin{\omega}$ & 0.16 $^{0.23}_{ - 0.28 }$ & \citet{Mills2019} \\
        Stellar reflex velocity, K$_\star$ (\ms)& 8.0$\pm$0.2  & This paper\footnotemark[3] \\
        Minimum mass, M$_\mathrm{pl}\,sini$ (\mjup) & 0.226$^{0.032}_{ - 0.031 }$  &  \citet{Mills2019}\\
        & & \\
        \hline
        \textbf{ Kepler-68}  & & \\
        $\bullet$       \underline{\textit{Stellar parameters}} &  & \\
        
        Spectral type & G1 & \citet{Grieves2018} \\
        Stellar mass, M$_\star$ (M$_\sun$) &  1.079$\pm$0.051     & \citet{Gilli2013}\\
        Stellar radius, R$_\star$ (R$_\sun$) & 1.243 $\pm$ 0.019 & \citet{Gilli2013}\\
        Stellar age, $\tau$ (Gyr) & 6.3 $\pm$ 1.7 & \citet{Gilli2013}\\
        Effective temperature, Teff (K) & 5793$\pm$74& \citet{Gilli2013}\\
        Metallicity (dex)&  0.12$\pm$0.074 [Fe/H] & \citet{Gilli2013}\\
        Surface gravity, log~g (log$_{10}$(\cms))&  4.281$\pm$0.008 & \citet{Gilli2013}\\
        Systemic velocity,  v$_\mathrm{sys}$ (\kms) & -20.762823 & DREAM I \\
        Limb-darkening coefficients $\mu_1$ & 0.148       & EXOFAST\footnotemark[1] \\
        $\,\,\,\,\,\,\,\,\,\,\,\,\,\,\,\,\,\,\,\,\,\,\,\,\,\,\,\,\,\,\,\,\,\,\,\,\,\,\,\,\,\,\,\,\,\,\,\,\,\,\,\,\,\,\,\,\,\,\,\,\,\,\,\,\,\,\,\,\,\,\,$ $\mu_2$  &   0.301        & EXOFAST\footnotemark[1] \\
        Stellar projected velocity, $v$sin$i_\star$ (\kms) & 0.5$\pm$0.5  & DREAM I\\
        Magnitude (J-band) & 8.975 $\pm$0.046  & \citet{Vizier}\\
        $\bullet$       \underline{\textit{Planetary parameters}} & & \\
        $\star$ Planet $b$  & & \\
        Orbital period, P (days) & 5.3987525913 $\pm$ 0.0000005231  & \citet{Gajdos2019}\\
        Transit epoch, T$_\mathrm{0}$ (BJD$_\mathrm{TDB}$) &   2455006.85878000 $\pm$  0.00007639  &\citet{Gajdos2019}\\\
        Eccentricity, e & 0.0 (fixed)&  \citet{Mills2019} \\
        Argument of periastron, $\omega$ (deg)& 90 (fixed)& \citet{Mills2019}\\
        Stellar reflex velocity, K$_\star$ & 2.70$^{+0.48}_{ - 0.49 }$ m s$^{-1}$ &  \citet{Mills2019} \\
        Scaled separation, a/R$_\star$ & 10.68 $\pm$  0.14 & \citet{Gilli2013}\\
        Orbital inclination, $i$ (deg) & 87.60 $\pm$  0.90&  \citet{Gilli2013}\\
        Planet-to-star radius, ratio R$_\mathrm{P}$/R$_\star$ & 0.01700 $\pm$  0.00046 &   \citet{Gilli2013}\\\
        Planetary mass, M$_\mathrm{pl}$ (\mjup) & 0.026 $^{+0.007}_{ -0.008}$&   \citet{Gilli2013} \\
        Planetary density, $\rho_\mathrm{pl}$ (\gcm) & 3.32 $^{+0.86}_{-0.98}$  & \citet{Gilli2013} \\
        Projected spin-orbit angle, $\lambda$ (deg) & non-detection & DREAM I \\
        Planet radial-velocity semi-amplitude, K$_\mathrm{pl}$(\kms) & 124.4$\pm$2.0 & This paper\footnotemark[2] \\
        Equilibrium temperature, T$_\mathrm{eq}$(K) &    1280$\pm$90 & \citet{Gilli2013} \\
        $\star$ Planet $c$  & & \\
        Orbital period, P (days) &  9.60502738150$\pm$ 0.0000132365 d&  \citet{Gajdos2019}\\
        Transit epoch, T$_\mathrm{0}$ (BJD$_\mathrm{TDB}$) &   2454969.38207000 $\pm$ 0.00110495 &  \citet{Gajdos2019}\\
        Eccentricity, e & 0.0 (fixed) & \citet{Mills2019} \\
        Argument of periastron, $\omega$ (deg) & 90 (fixed) & \citet{Mills2019} \\
        Stellar reflex velocity, K$_\star$ (\ms) & 0.59 $^{+0.50}_{ -0.52}$ &  \citet{Mills2019}\ \\
        
        $\star$ Planet $d$  & & \\
        Orbital period, P (days) & 634.6 $^{+4.1}_{ -3.7 }$ &   \citet{Mills2019}\ \\
        Transit epoch, T$_\mathrm{0}$ (BJD$_\mathrm{TDB}$) & 2455878 $\pm$ 11   &   \citet{Mills2019}\ \\
        Eccentricity, e & 0.112 $_{ -0.034}^{+0.035}$ & \citet{Mills2019}\ \\
        Argument of periastron, $\omega$ (deg) & -64.74$_{-20.63}^{+25.78}$  &  \citet{Mills2019}\ \\
        Stellar reflex velocity, K$_\star$ (\ms) & 17.75$^{+0.50}_{-0.49}$ &  \citet{Mills2019}\ \\
        & & \\
        \hline
        \textbf{ WASP-47}  & & \\
        
        $\bullet$       \underline{\textit{Stellar parameters}} &  & \\

        Spectral type & G9 & \citet{Hellier2012}\\
        Stellar mass,   M$_\star$ (M$_\sun$)&  1.040 $\pm$0.031  & \citet{Vanderburg2017}\\
        Stellar radius,   R$_\star$ (R$_\sun$)& 1.137 $\pm$ 0.013  & \citet{Vanderburg2017}\\
        Stellar age, $\tau$ (Gyr) & 6.5 $^{+2.6}_{-1.2}$ & \citet{almenara2016}\\
        Effective temperature, Teff  (K) & 5552$\pm$75 & \citet{Vanderburg2017}\\
        Metallicity (dex)& 0.38$\pm$0.05 [Fe/H]  & \citet{Vanderburg2017}\\
        Surface gravity, log~g (log$_{10}$(\cms))&   4.3437$\pm$0.0063 & \citet{Vanderburg2017}\\
        Systemic velocity,  v$_\mathrm{sys}$ (\kms) & -25.847809453 & \citet{Vanderburg2017}\\
        Limb-darkening coefficients $\mu_1$& 0.179      & EXOFAST\footnotemark[1] \\
        $\,\,\,\,\,\,\,\,\,\,\,\,\,\,\,\,\,\,\,\,\,\,\,\,\,\,\,\,\,\,\,\,\,\,\,\,\,\,\,\,\,\,\,\,\,\,\,\,\,\,\,\,\,\,\,\,\,\,\,\,\,\,\,\,\,\,\,\,\,\,\,$ $\mu_2$  &   0.299        & EXOFAST\footnotemark[1] \\
        Stellar projected velocity, $v$sin$i_\star$ (\kms) & 1.80$^{+0.24}_{-0.16}$  & DREAM I\\
        Magnitude (J-band) & 10.613$\pm$0.022  & \citet{Vizier}\\ 
        $\bullet$       \underline{\textit{Planetary parameters}} & & \\
        $\star$ Planet $b$  & & \\
        Orbital period, P (days) &  4.1591492 $\pm$0.000006 & \citet{Bryant2022}\\
        Transit epoch, T$_\mathrm{0}$ (BJD$_\mathrm{TDB}$) &  2457007.932103 $\pm$ 0.000019 & \citet{Bryant2022}\\
        Eccentricity, e & 0 (fixed)& \citet{Bryant2022}\\
        Argument of periastron, $\omega$ (deg) &  90 (fixed)& \citet{Bryant2022}\\
        Stellar reflex velocity, K$_\star$ (\ms) &  140.84 $\pm$0.40 & \citet{Bryant2022} \\
        $\star$ Planet $c$  & & \\
        Orbital period, P (days) & 588.8 $\pm$ 2.0 & \citet{Bryant2022}\\
        Transit epoch, T$_\mathrm{0}$ (BJD$_\mathrm{TDB}$) & 2457763.1 $\pm$ 4.3 & \citet{Bryant2022}\\
        Eccentricity, e & 0.296 $\pm$ 0.016 & \citet{Bryant2022}\\
        Argument of periastron, $\omega$ (deg) & 112. $\pm$ 4.3  & \citet{Bryant2022}\\
        Stellar reflex velocity, K$_\star$ (\ms) &  31.04 $\pm$ 0.40 & \citet{Bryant2022} \\
        
        $\star$ Planet $d$  & & \\
        Orbital period, P (days) & 9.03052118 $\pm$0.00000753  & DREAM I \\
        Transit epoch, T$_\mathrm{0}$ (BJD$_\mathrm{TDB}$) & 2459426.5437 $\pm$0.0028    & DREAM I \\
        Eccentricity, e & 0.010$^{+0.011}_{-0.007}$&  \citet{Bryant2022}\\
        Argument of periastron, $\omega$ (deg) & 16.5 $^{+84.2}_{-98.6}$ & \citet{Bryant2022} \\
        Stellar reflex velocity, K$_\star$  (\ms)& 4.26 $\pm$0.37 & \citet{Bryant2022} \\
        Scaled separation, a/R$_\star$ & 16.34 $^{+0.08}_{-0.11}$& \citet{Bryant2022}\\
        Orbital inclination, $i$ (deg) & 89.55  $^{+0.30}_{-0.27}$& \citet{Bryant2022}\\
        Planet-to-star radius ratio, R$_\mathrm{P}$/R$_\star$ & 0.02876 $\pm$ 0.00017&  \citet{Bryant2022}\\
        Planet mass, M$_\mathrm{pl}$ (\mearth) & 14.2$\pm$1.3 & \citet{Bryant2022}\\
        Planetary density, $\rho_\mathrm{pl}$ (\gcm)  &  1.72$\pm$0.17  &  \citet{Bryant2022}\\
        Projected spin-orbit angle, $\lambda$ (deg)& 0$\pm$24  & DREAM I \\
        Planet radial-velocity semi-amplitude, K$_\mathrm{pl}$(\kms) & 103.6$\pm$1.0 & This paper\footnotemark[2] \\
        Equilibrium temperature, T$_\mathrm{eq}$(K) & 919$\pm$13  & This paper\footnotemark[4] \\
        
        $\star$ Planet $e$  & & \\
        Orbital period, P (days) &  0.7895933 $\pm$ 0.0000044& \citet{Bryant2022}\\
        Transit epoch, T$_\mathrm{0}$ (BJD$_\mathrm{TDB}$) & 2457011.34862  $\pm$ 0.00030& \citet{Bryant2022}\\
        Eccentricity, e &0 (fixed)& \citet{Bryant2022}\\
        Argument of periastron, $\omega$ (deg) & 90 (fixed) & \citet{Bryant2022}\\
        Stellar reflex velocity, K$_\star$  (\ms)& 4.55  $\pm$  0.37 & \citet{Bryant2022} \\               
        \label{Tab_parameters}
\end{longtable}
\tablefoot{     
        \begin{enumerate}
\item For a homogeneous analysis we used quadratic limb-darkening coefficients derived using the EXOFAST calculator \url{https://astroutils.astronomy.osu.edu/exofast/limbdark.shtml}
        \citep{exofast} in the J-band.
\item K$_\mathrm{pl}=\frac{2\pi a}{P}\frac{\sin{i}}{\sqrt{1-e^2}}=(\frac{2\pi G}{P})^{\frac{1}{3}}\frac{(M_\star+M_\mathrm{pl})^{\frac{1}{3}}\sin{i}}{\sqrt{1-e^2}}$.
\item K$_\mathrm{\star}=(\frac{2\pi G}{P})^{\frac{1}{3}}\frac{M_\mathrm{pl}*\sin{i}}{M_\star^{2/3}\sqrt{1-e^2}}$.
\item $\rm T_{\rm {eq}}=\rm T_{\star} \left( \frac{ \rm R_{\star}}{2\,a} \right)^{1/2} (1-\rm{A})^{1/4}$, where $R_{\star}$ is the stellar radius, $a$ is the semi-major axis, $A$ is the geometric albedo, we assumed an albedo of 0.2 \citep{CrossfieldTrends}.
\item Derived from \url{http://www.pas.rochester.edu/~emamajek/EEM_dwarf_UBVIJHK_colors_Teff.txt}
\end{enumerate}}

        \twocolumn
        
        \begin{figure*}
                \centering
                \includegraphics[width=0.8\linewidth,height=16cm]{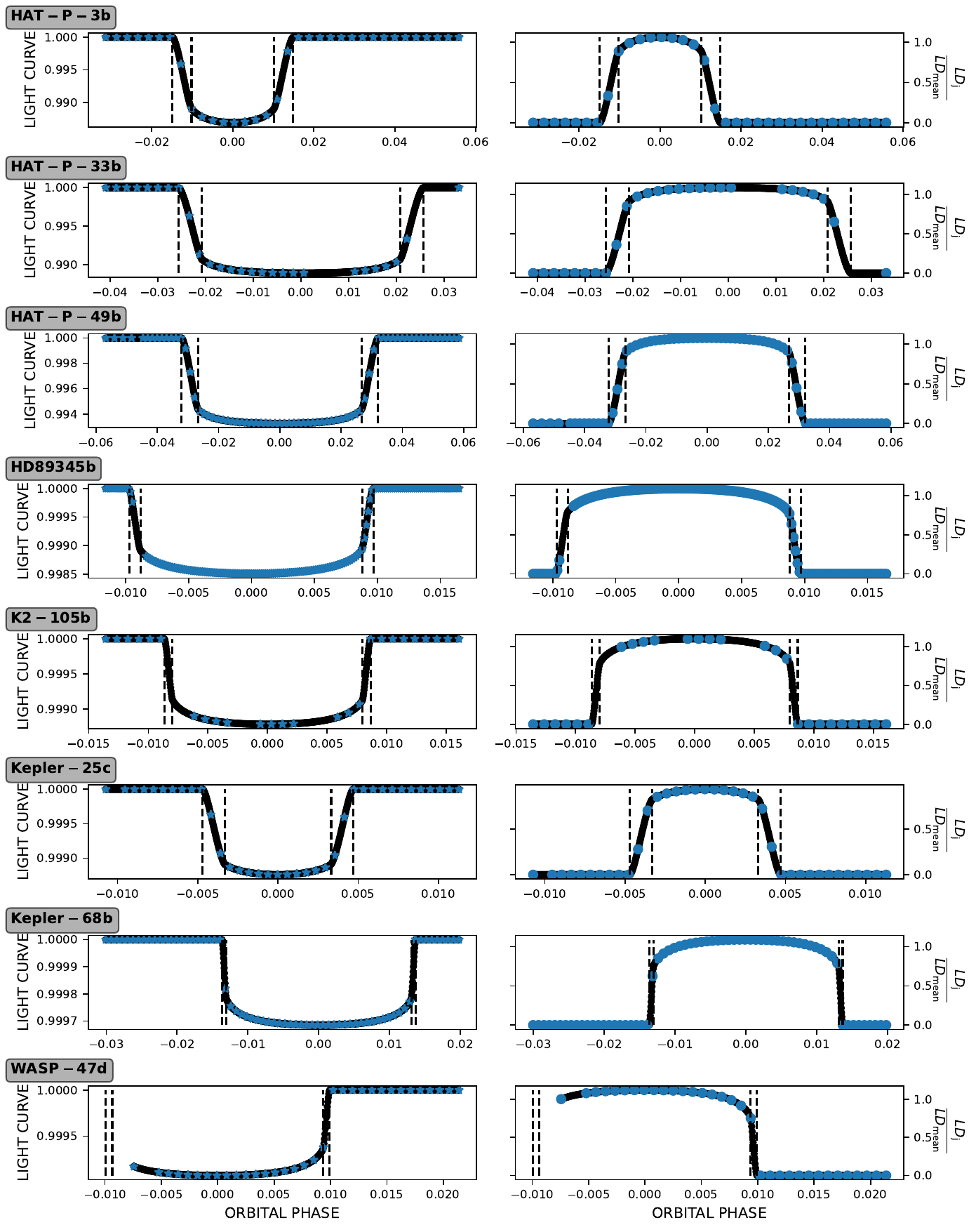}
                \caption{Light curves (left panels) and the limb-darkening ratio, namely, $\frac{LD_\mathrm{j}}{LD_\mathrm{mean}}$ with $j$ the orbital phase (right panels) computed with the Python $batman$ code  \citep{batman}  and  the  system  parameters  from  Table~\ref{Tab_parameters}. The contact points t$_1$, t$_2$, t$_3$, and t$_4$ are marked with vertical black lines. Blue dots are the light curves and limb-darkening ratio at the observed phases.}
                \label{LC_panels}
        \end{figure*}

        \begin{table*}[]
                \centering
                \caption{Initial parameters adopted for Molecfit.}
                \label{tab_molec}
                \begin{tabular}{c|c c}
                        \hline \hline
                        Parameter & Value & Significance \\
                        \hline
                        ftol & 10$^{-5}$  & $\chi^2$ convergence criterion \\
                        xtol & 10$^{-5}$ & Parameter convergence criterion \\
                        molecules & H$_2$O \\
                        ncont & 4 & Degree of coefficient for continuum fit \\
                        a$_0$ & 1 & Constant term for continuum fit \\
                        $\omega_\mathrm{gaussian}$ & 2.25 & FWHM of Gaussian in pixels \\
                        kernel size & 3 & Size of Gaussian kernel in FWHM \\
                        slit width & 0.5 arcsec & Slit width \\ 
                        MIPAS profile & equ & Equatorial profile \\
                        Atmospheric profile & True & Fixed grid \\
                        PWV & -1 & Input for water vapor profile \\

                \end{tabular}
        \end{table*}

        \begin{figure*}
                \centering
                \includegraphics[width=\linewidth,height=9cm]{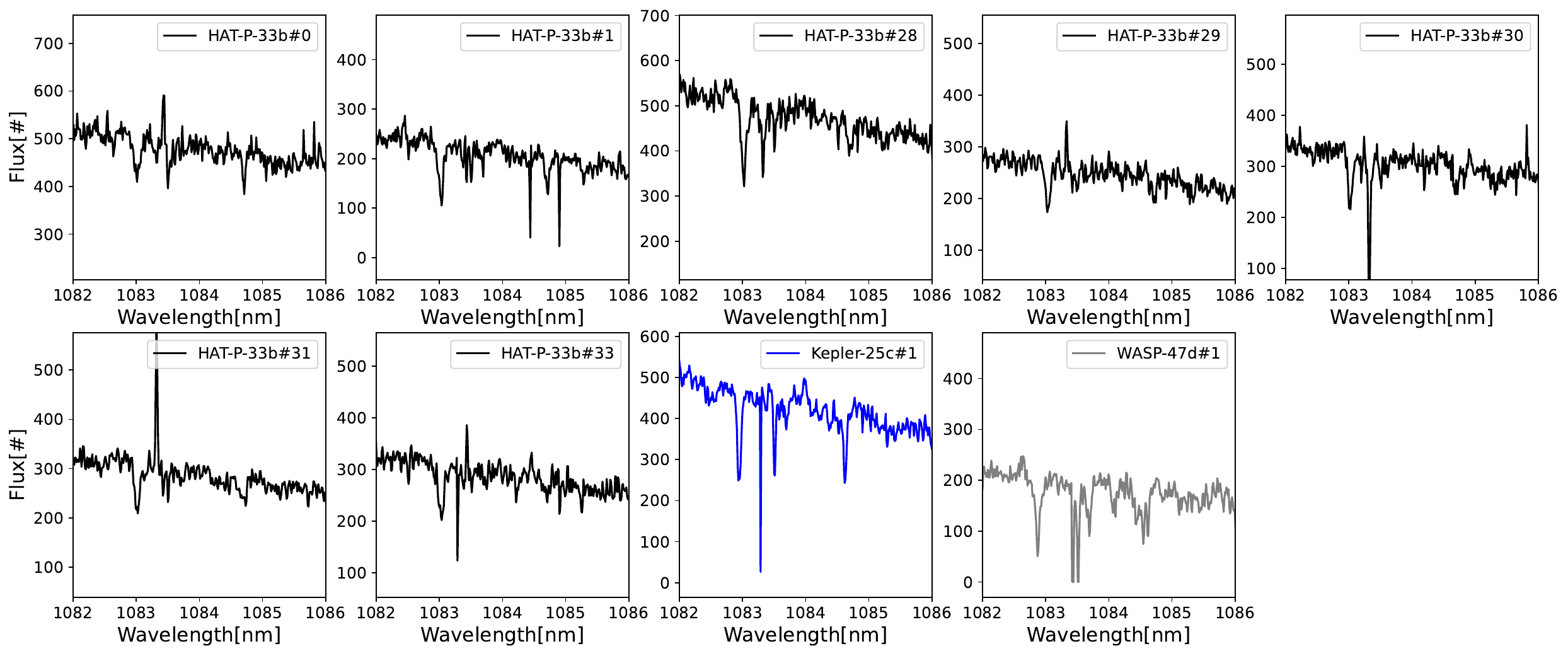}
                \caption{Exposures not considered in the analysis for the presence of outliers or low counts. Different colors correspond to different observing nights}.
                \label{App_bad}
        \end{figure*}

        \begin{figure*}
                \centering
                \includegraphics[width=\linewidth,height=8.cm]{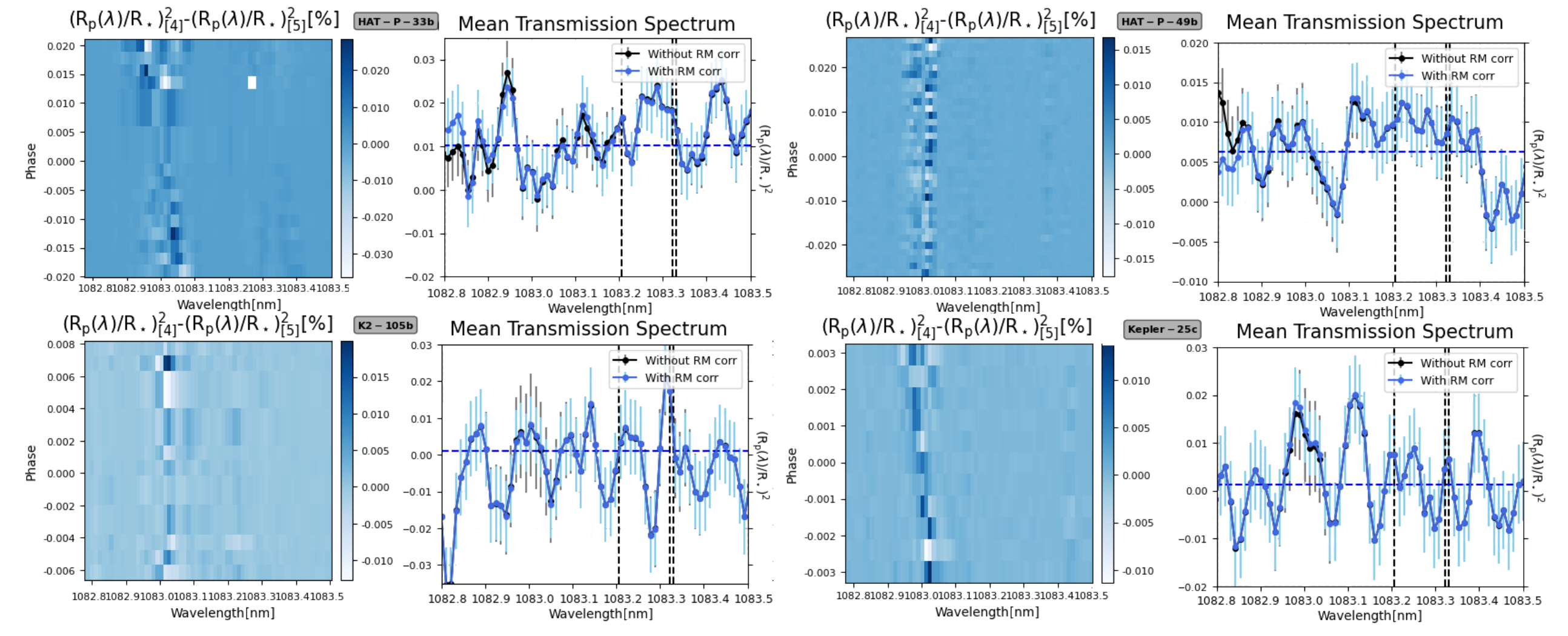}
                \caption{Difference between the full in-transit transmission spectra in stellar rest frame when both accounting and not accounting for the RM effects (left panel). Average full in-transit transmission spectrum in the planet rest frame when both accounting (blue line) and not accounting (black line) for the RM effect (right panel). Left panels also show the difference between the transmission spectra calculated with formula \ref{eq_T}, with the RM correction, and \ref{eq_T2}, no RM correction applied, in the stellar rest frame. The RM signal creates a signature that is comparable to the dispersion in a given transmission spectrum and furthermore smoothes out when averaging over the transit because the RM signature shifts along the transit chord radial velocities.}
                \label{ros_dif}
        \end{figure*}
        
        \begin{figure*}
                \centering
                \includegraphics[width=\linewidth]{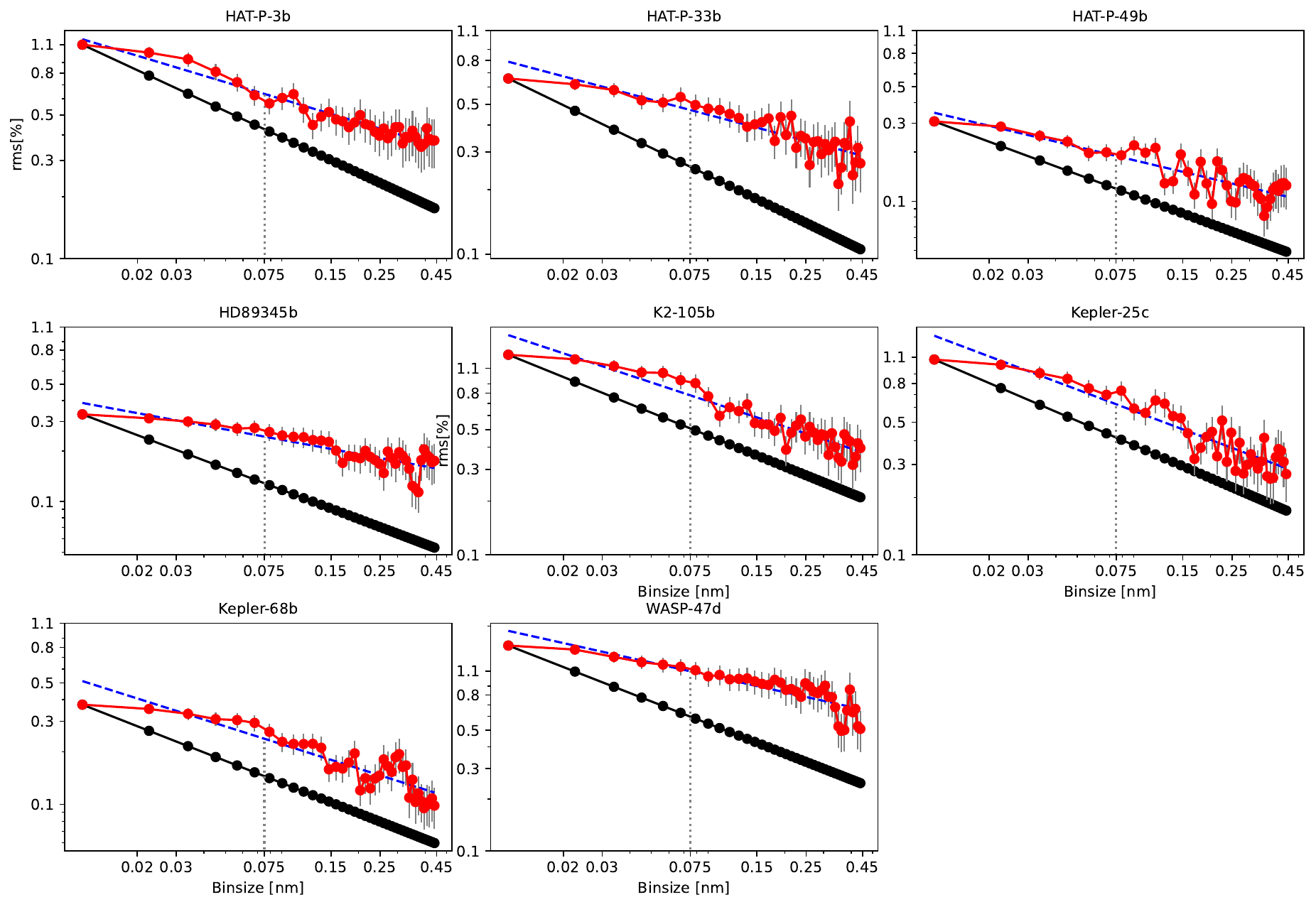}
                \caption{Allan plot computed on the average transmission spectra. The back lines are the expected rms for white noise (which scales with the number of points for bin).  The red dotted curves are the standard deviation of the transmission spectrum after various binning of different bin size. The dashed blue
                        lines are the best fit for the red curves (computed in log-log scale). The vertical gray lines are the derived 1$\sigma$ uncertainty at 0.075~nm. The error bars denote this 1$\sigma$uncertainty of the rms \citep{Cubillos2017}.}
                \label{Allan}
        \end{figure*}

        \begin{figure*}[h]
                \centering
                \includegraphics[width=\linewidth]{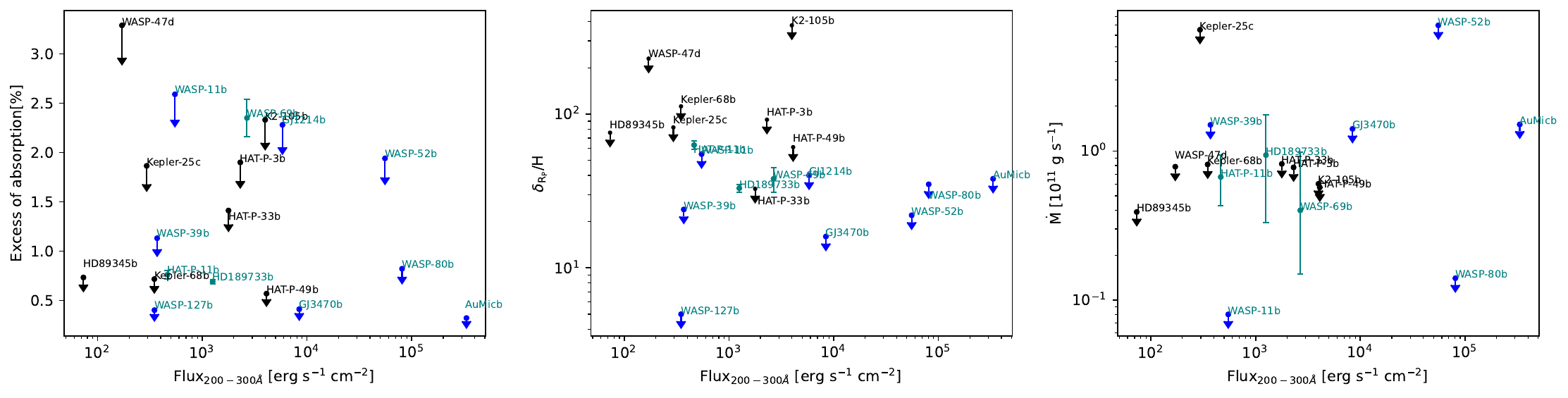}
                \caption{Correlation plots with the insolation level of mid-UV flux (200-300 \AA). Same plot as in Fig.~\ref{fig_result}. As for the XUV flux between 5 and 504 \AA, the \ion{He}{I}
                        absorption signal correlates with the received mid-UV flux; whereas when they are translated in terms of mass-loss rates, these trends seem to disappear. }
                \label{mid_flux}
        \end{figure*}
        
        \clearpage
        
        \section{Gaussian processes to derive upper limits on the helium absorption} \label{GPL_sec}
        To have a better description of the correlated noise present in the data, we performed a Differential Evolution Markov chain Monte Carlo (DEMCMC) fit of a Gaussian profile fixing the position at  1083.326 nm and the FWHM at 0.07 nm and varying the peak value, an offset for the continuum, an uncorrelated jitter, and a correlated noise modeled with a Gaussian process (GP) and a squared exponential kernel. From the posterior distribution, we were therefore able to derive the 3$\sigma$ upper limits  (the value to which 95\% of the peak distribution is subject) at the position of the helium triplet marginalized over an uncorrelated jitter and the presence of correlated noise. The values are reported in Table~\ref{GPL}.

        \begin{table}[h]
                \centering
                \caption{Upper limits on the excess of absorption.}
                \label{GPL}
                \begin{tabular}{c|c }
                        \hline \hline
                        Parameter & 3$\sigma$[\%] \\
                        \hline
                        HAT-P-3b    &   1.48\\
                        HAT-P-33b    &   1.17\\
                        HAT-P-49b   &   0.95\\
                        HD89345b    &  0.63\\
                        K2-105b     &  2.57\\
                        Kepler-25c  &    1.29\\
                        Kepler-68b  &    0.64\\
                        WASP-47d    &   3.11\\         
                \end{tabular}
\tablefoot{     3$\sigma$ upper limits calculated with the Gaussian processes.}
        \end{table}

\end{appendix}

\end{document}